\documentclass[review,12pt]{elsarticle}

\usepackage{setspace}
\usepackage{subcaption}
\usepackage{multirow}
\usepackage{multicol}
\usepackage{graphicx}
\RequirePackage{amsmath}
\usepackage{bm}
\usepackage{xfrac}
\usepackage{float}
\usepackage{arydshln} 

\usepackage{amsmath,esint,amsfonts}

\usepackage[most]{tcolorbox}

\usepackage{lineno,hyperref}
\usepackage{cleveref} 

\journal{International Communications in Heat and Mass Transfer}

\biboptions{sort&compress} 

\begin{document}

\begin{frontmatter}

\title{Numerical Simulations of Frost Growth Using Mixture Model on Surfaces with Different Wettability}

\author{Shantanu Shahane}
\author{Yuchen Shen}
\author{Sophie Wang\fnref{Corresponding Author}}
\address{Department of Mechanical Science and Engineering\\
	University of Illinois at Urbana-Champaign \\
	Urbana, Illinois 61801}
\fntext[Corresponding Author]{Corresponding Author Email: \url{wangxf@illinois.edu}}


%

\begin{abstract}
Frost growth on cold surfaces is a transient process with coupled heat and mass transfer. Due to multiple factors such as humidity, temperature, flow velocity and constantly changing thermal properties as frost grows, precise prediction can be challenging. Especially when the geometry of the frosting surfaces gets complicated, it requires a balance of computing accuracy and efficiency. In this work, a numerical model is developed to predict frost growth considering the effect of the above parameters. Mixture model is adapted to improve the computational efficiency and the unstructured grids add the flexibility to extend the model to complex geometries. The predicted frost growth rate matches well with the experimental data reported in the literature under similar conditions. The model predicts reasonable growth trend of frost as the surface temperature, air temperature, humidity and flow velocity vary. The surface wettability effect is well captured at the early stage of frosting and it shows a higher frost growth rate on surfaces with a higher wettability.
\end{abstract}


\end{frontmatter}
\section{Introduction}
Frost buildup on surfaces is undesirable in a lot of engineering applications, such as frost (ice in some cases) on aircraft wings, electrical transition lines, traffic surfaces etc. In air source heat pumping or refrigeration system, a layer of frost on fins of heat exchanger can increase thermal resistance and block air passage, which results in low heat transfer and high pumping power, ultimately a low system COP.
\par Frost, which is a composition of ice crystal and air in a porous structure, has variable thermal properties (density, thermal conductivity etc.) depending on the conditions it subjects to and the history it goes through \cite{shen2019real,shen2020condensation1}. When the surrounding air temperature is below dew point and the surface temperature is subzero, water vapor in the air first condenses on the cold surfaces and then becomes solidified, which is called condensation-frosting. Under extremely low air temperature, desublimation frosting can occur. As frost growing, the local temperature of frost, humidity of the trapped air varies and ultimately affects the following frost growth. So frosting is a highly transient coupled heat-mass transfer process, and multiple factors including surface temperature, wettability, roughness \cite{shen2020condensation2, kandula2011frost, hermes2009study, yun2002modeling, rabbi2021wettability, wang2016design, liu2016strategies, chu2018frost}, air temperature, humidity and velocity \cite{cheng2003observations, lee2003prediction, niroomand2019experimental, chen2019simulation} can affect the process.
\par Here, we summarize the modeling approaches in some important research papers published in the recent years. \citet{lee2003prediction}, \citet{yang2005modeling} and \citet{chen2019simulation} developed two dimensional models for simulating frost growth over flat plate. The models solve separate sets of laminar steady state momentum, energy and mass conservation equations over  regions of frost and humid air. Interface condition is applied over the frost surface. \citet{cui2011new_ate} and \citet{wu2016phase} also simulated frost growth over a two dimensional flat surface using the Eulerian--Eulerian two phase model for the system consisting humid air and ice droplets as primary and secondary components respectively. \citet{wu2017frosting} further improved the model by adding a non-dimensional phase change driving force based on Gibbs free energy. \citet{cui2011new_ijhff} used a similar strategy to model coupled heat transfer and frost growth on pin--and--tube heat exchanger surfaces. A one dimensional model was developed by \citet{kandula2011frost}, \citet{loyola2014modeling} and \citet{el2014mathematical} to estimate the rate of frost growth due to laminar flow over flat surfaces. They assumed that the frost grows normal to the plate and hence, all the variations are modeled only in the normal direction. \citet{kim2015frosting} used a single domain Eulerian two phase model to simulate frost growth over flat plates in two dimensions. They distinguished the air and ice phases by volume fraction. \citet{armengol2016modeling} modeled frost growth on the fin--and--tube heat exchanger using a two dimensional domain. Laminar momentum, mass and energy conservation equations are solved in the air subdomain whereas, only energy and densification equations are solved in the frost subdomain.
\par Some of the above described papers model frost growth in one dimensional settings normal to the plate. Thus, they do not consider the effect of gradients in the direction of air flow. The remaining two dimensional models either use a two domain or an Eulerian multiphase approach. The two domain method needs interface condition at the air--frost boundary. A multiphase approach on the other hand requires the source terms which couple the phases. An accurate estimation of interface condition or source terms is a challenge for problems over complex domains having nontrivial interfaces. Hence, in this work, we have developed an approach based on mixture model which uses a single set of conservation equations with a volume fraction to distinguish between the phases. Mixture model has been used extensively in the study of phase change systems \cite{shahane2019finite, shahane2019numerical, bennon1987continuum, plotkowski2015estimation, voller1987fixed, bartrons2019fixed}. Recently, \citet{bartrons2019fixed} used a mixture model with frost as a porous structure to analyze frost growth over a flat plate in two dimensions with a two step predictor--corrector approach. We use the semi--implicit pressure splitting method to solve the momentum equations coupled with mass conservation of water and energy equations. The governing equations are discretized on unstructured grids and thus, frost growth can be simulated on complex geometries in two and three dimensions. The ease of implementation and computational efficiency makes this an attractive approach compared to the traditional two domain or Eulerian multiphase methods.
\par In recent decades, it has been reported that the frosting behavior is affected by surface wettability \cite{yue2018freezing, el2014effect, huang2011preparation, liu2008frost, sommers2018role, cai2011study, wang2015effects, sommers2016densification}. Majority of these studies concluded that the surface wettability mainly contributed to the initial stage of the condensation frosting \cite{yue2018freezing, el2014effect, huang2011preparation, liu2008frost, sommers2018role}, such as altering the profile of initial frost layer, which only consists of ice droplets and surrounding surface. Frost growth on ice droplets over the initial layer is still an area of active research. Experimental observation shows that the ice crystal preferably grows on the ice droplet rather than the substrate surface \cite{liu2017distinct}, which can be well captured by Classic Nucleation Theory (CNT). CNT was first proposed by \citet{volmer1926nucleus} and has been commonly used for prediction of nucleation rate since it has shown agreement with experimental data in many studies. \citet{wolk2001homogeneous} compared the measured homogeneous nucleation rate of water with CNT developed by \citet{becker1935kinetische}. The result shows a good agreement between measurements and predictions when the temperature is between 220 K to 260 K for both light and heavy water. Other than the homogeneous nucleation such as cloud formation, CNT can also be applied for heterogeneous nucleation with the usage of Volmer theory \cite{volmer1939kinetik}. \citet{twomey1959experimental} computed the critical supersaturation for heterogeneous nucleation for given nucleation rate and compared with observations. The dependence of critical supersaturation was well predicted for a large range of surface contact angles. \citet{xu2015heterogeneous} used CNT to analyze the nucleation rate for surfaces with different wettability and microstructures. Their modeling result shows that the nucleation rate is higher on a hydrophilic surface than that on a hydrophobic surface with any microstructures. Based on CNT, \citet{varanasi2009spatial} achieved a spatial control of the heterogeneous nucleation for water by generating a wettability gradient on the surface. The nucleation rate on a hydrophilic surface can be $\sim 10^{129}$ times larger than hydrophobic surface, according to their estimate. All these previous studies show that CNT is an appropriate method in analyzing the nucleation related problem. As for the initial frost layer, it can be viewed as a biphilic surface, containing hydrophobic part (surface without ice drops) and hydrophilic part (ice drops). We therefore believe that the following crystal growth on initial frost layer can be predicted using CNT.

\begin{table}[t]
	Nomenclature
\begin{center}
	\begin{tabular}{|@{}p{0.08\textwidth}@{} @{}p{0.42\textwidth}@{}| @{}p{0.08\textwidth}@{} @{}p{0.42\textwidth}@{}|}
		\hline & & &\\
		\hspace{0.1cm} $T$ & Temperature & 	\hspace{0.1cm} $p$ & Pressure\\
		\hspace{0.1cm} $\bm{u}$ & Velocity vector $(u,v)$ & 	\hspace{0.1cm} & \\
		\hspace{0.1cm} $\alpha_{ice}$ & Ice volume fraction & 	\hspace{0.1cm} $\alpha_{ha}$ & Humid air volume fraction\\
		\hspace{0.1cm} $\mu$ & Dynamic viscosity & 	\hspace{0.1cm} $Y$ & Humidity mass fraction\\
		\hspace{0.1cm} $M$ & Molar mass & 	\hspace{0.1cm}  $R$ & Ideal gas constant \\
		\hspace{0.1cm} $ha$ & Humid air (subscript) & 	\hspace{0.1cm} $da$ & Dry air (subscript)\\
		\hspace{0.1cm} $wv$ & Water vapor (subscript) & 	\hspace{0.1cm} $m$  & Mixture (subscript)\\
		\hline
	\end{tabular}
\end{center}
\end{table}
\section{The Mixture Model}
\begin{figure}[H]
	\centering
	\includegraphics[width=0.8\textwidth]{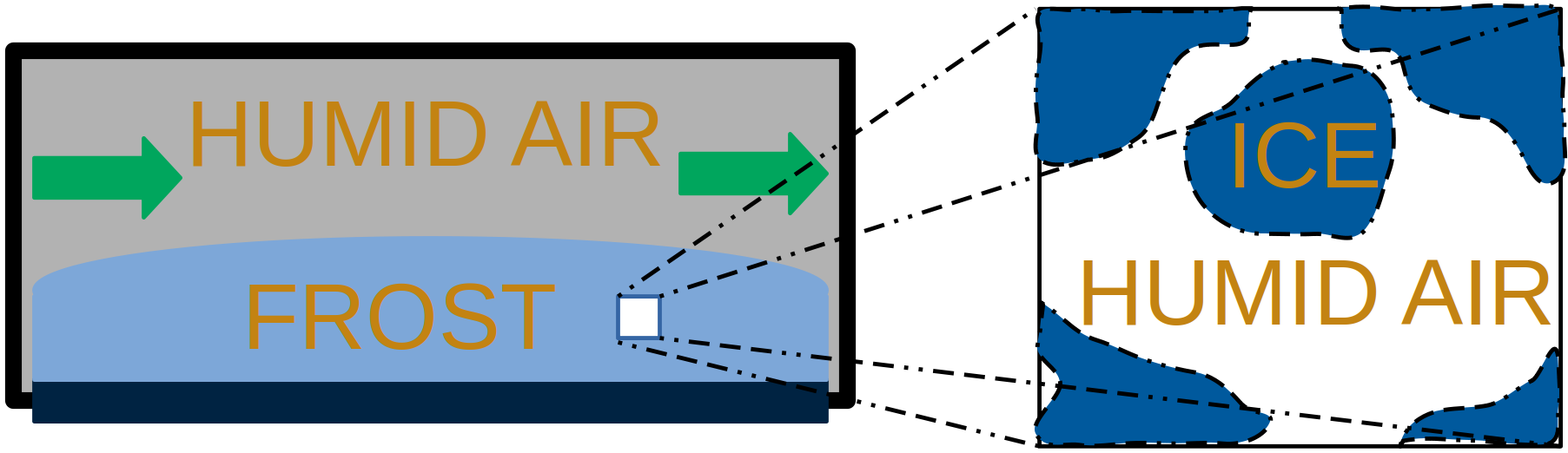}
	\caption{Schematic of Mixture Model}
	\label{Fig:Schematic of Mixture Model}
\end{figure}

\Cref{Fig:Schematic of Mixture Model} shows schematic of a domain consisting of cold plate at the bottom exposed to humid air flowing from left to right. Humid air cools down after coming in contact with the surface at sub--zero temperature and its capacity to hold water vapor drops. Hence, some portion of the vapor condenses as liquid or desublimates as frost depending on the conditions. Frost is a porous mixture of ice and air as shown in the zoomed image. In the mixture model, a single set of field variables (velocity, pressure, temperature etc.) are defined for the entire domain. Volume fraction is used to distinguish between multiple components in this case, humid air and frost. Benefit of this method over other multiphase approaches is that we do not need to solve separate conservation equations for each component. Moreover, due to definition of volume fraction, explicit interface tracking and interface conditions are not required. Hence, this approach is more tractable, easy to implement and computationally efficient. In this section, we derive all the equations for two dimensional domains for the sake of brevity. However, this method can be extended for three dimensional problems as well.
\subsection{Governing Equations of Conservation}
Conservation of mass for humid air is written as follows:
\begin{equation}
	\nabla \bullet (\rho_{ha} \alpha_{ha} \bm{u}) = \frac{\partial (\rho_{ha} \alpha_{ha} u)}{\partial x} + \frac{\partial (\rho_{ha} \alpha_{ha} v)}{\partial y} =0
	\label{Eq:continuity}
\end{equation}
where, $\alpha_{ha}$ denotes the volume fraction of humid air in a given control volume. Volume fractions of ice and humid air should sum to unity: $\alpha_{ice} + \alpha_{ha} = 1$. We assume incompressibility of humid air and neglect the term having time derivative of density \cite{bartrons2019fixed}. The deposited frost has zero velocity. Hence, the conservation of momentum in $X$ and $Y$ directions is formulated only for humid air as shown in \cref{Eq:momentum x,Eq:momentum y}. The mixture model equations are formulated using the macroscopic density instead of material or microscopic density. Hence, the unsteady and convection terms in all the conservation equations (\cref{Eq:continuity,Eq:momentum x,Eq:momentum y,Eq:water conservation,Eq:enthalpy conservation}) are scaled by the volume fraction of humid air ($\alpha_{ha}$) which may be thought as a porosity or void fraction. Such an approach is suggested by \citet{harlow1975numerical}.
\begin{equation}
	\frac{\partial (\rho_{ha} \alpha_{ha} u)}{\partial t} + \nabla \bullet (\rho_{ha} \alpha_{ha} \bm{u} u) =
	-\frac{\partial p}{\partial x} +  \nabla \bullet (\mu_{ha} \nabla u) - K_d u
	\label{Eq:momentum x}
\end{equation}
\begin{equation}
	\frac{\partial (\rho_{ha} \alpha_{ha} v)}{\partial t} + \nabla \bullet (\rho_{ha} \alpha_{ha} \bm{u} v) =
	-\frac{\partial p}{\partial y} +  \nabla \bullet (\mu_{ha} \nabla v) - K_d v
	\label{Eq:momentum y}
\end{equation}
where, $K_d$ is the coefficient of the Darcy drag term estimated empirically \cite{voller1987fixed}:
\begin{equation}
	K_d = \frac{C_0 \alpha_{ice}^2}{(1-\alpha_{ice})^3 + \epsilon_0}
	\label{Eq:Darcy K}
\end{equation}
$C_0$ is an arbitrary constant which is set to 1.5E6 \cite{bartrons2019fixed} in this work and $\epsilon_0$ is set to a small value (such as 1E-12) to avoid division by zero in the case of $\alpha_{ice}=1$. If a given control volume has no ice ($\alpha_{ice}=0$), the coefficient $K_d$ takes a value of zero. Thus, the Darcy drag term in the momentum equations becomes zero. On the other hand, when the control volume has some frost deposited ($0<\alpha_{ice}\leq 1$), $K_d$ takes a positive value which is added to the diagonal of the discretized and linearized momentum equations (\cref{Sec:Pressure--Velocity Coupling}). This acts as a drag to the velocities. Hence, in the regions with frost, the velocity of humid air drops corresponding to the porosity. Similar approach is used by several researchers to model phase change systems \cite{shahane2019finite, bartrons2019fixed, plotkowski2015estimation}.
\par We define humidity ratio as mass of water vapor per unit mass of humid air: $Y={m_{wv}}/{(m_{wv}+m_{da})}$. Let $M_{wv}$, $M_{da}$ and $M_{ha}$ denote molecular masses of water vapor, dry air and humid air respectively. Molecular mass of humid air:
\begin{equation}
	M_{ha}= \frac{m_{wv}+m_{da}}{\frac{m_{wv}}{M_{wv}} + \frac{m_{da}}{M_{da}}}  = \left[\frac{Y}{M_{wv}} + \frac{1-Y}{M_{da}}\right]^{-1}
\end{equation}
Density of humid air using the ideal gas law:
\begin{equation}
	\rho_{ha}= \frac{P M_{ha}}{R T}
	= \frac{P \left(\frac{M_{ha}}{M_{da}}\right)}{R_{da} T}
	= \frac{P}{R_{da} T}\left(\frac{1}{Y\frac{M_{da}}{M_{wv}} + 1-Y}\right)
\end{equation}
At the same pressure, density of dry air is: $\rho_{da}=\frac{P}{R_{da} T}$. Substituting in above equation gives:
\begin{equation}
	\rho_{ha} = \rho_{da} \left(\frac{1}{Y\frac{M_{da}}{M_{wv}} + 1-Y}\right)
	\label{Eq:humid air density}
\end{equation}
Any given control volume may consist of water vapor, dry air and ice. Hence, the mixture density can be estimated as follows:
\begin{equation}
	\rho_m = \alpha_{ice} \rho_{ice} + \alpha_{ha} \rho_{ha}
	\label{Eq:mean density}
\end{equation}
Water vapor in air partially desublimates to frost. Hence, we formulate conservation of water:
\begin{equation}
	\frac{\partial (\rho_{ha} Y + \rho_{ice} \alpha_{ice})}{\partial t} + \nabla \bullet \left(Y \rho_{ha} \vec{u} \alpha_{ha}\right) = \nabla \bullet \left(\tau D_{wv} \rho_{ha} \nabla Y \right)
	\label{Eq:water conservation}
\end{equation}
where, $\tau$ is the diffusion resistance factor and $D_{wv}$ is the diffusivity of water vapor in air. We use the model of \citet{le1997modelling} to estimate the diffusion resistance factor:
\begin{equation}
	\tau = \frac{\alpha_{ha}}{1-0.58\alpha_{ice}} + 50 \alpha_{ice} \alpha_{ha}^{10}
	\label{Eq:diffusion resistance}
\end{equation}
The first term in \cref{Eq:water conservation} models conversion of water vapor in humid air into ice. Since ice does not move, only water vapor in air undergoes convection and diffusion. Conservation of energy is written as:
\begin{equation}
	\frac{\partial \rho_m h_m}{\partial t} + \nabla \bullet \left(h_{ha} \rho_{ha} \vec{u} \alpha_{ha}\right) = \nabla \bullet \left(k_m \nabla T \right)+ \nabla \bullet \left((h_{wv}-h_{da})\tau D_{wv} \rho_{ha} \nabla Y \right)
	\label{Eq:enthalpy conservation}
\end{equation}
The last term in the \cref{Eq:enthalpy conservation} estimates the change of enthalpy due to diffusion of water vapor in air. This term is modeled implicitly by expressing the enthalpy of water vapor and dry air in terms of temperature and specific heat capacity. The enthalpy is expressed using the specific heat capacity: $h_{da}= C_{p_{da}} T$, $h_{wv}= C_{p_{wv}} T$ and $h_{ice}= C_{p_{ice}} T - L_{sub}$ where, $L_{sub}$ denotes the latent heat of sublimation.
The specific heat capacity of dry air and water vapor is estimated using best fit polynomial expressions as a function of temperature \cite{bartrons2018finite, eckertanalysis, lide1994crc}. The specific heat capacity of humid air and frost, in \cref{Eq:specific heat capacities}, is defined as weighted sums of individual components \cite{bartrons2018finite, kandula2011frost}.
\begin{equation}
	C_{p_{ha}} = Y C_{p_{wv}} + (1-Y) C_{p_{da}} \text{ and } C_{p_{m}} = \alpha_{ice} C_{p_{ice}} + \alpha_{ha} C_{p_{ha}}
	\label{Eq:specific heat capacities}
\end{equation}
Thermal conductivity of mixture is computed using the Studnikov relation \cite{fessler1979wetair, studnikov1970viscosity} for humid air and Na--Webb relation \cite{na2004new} for frost:
\begin{equation}
	k_m=
	\begin{cases}
		(k_{wv} \widetilde{m_{wv}} + k_{da} \widetilde{m_{da}}) \left(1+ \frac{\widetilde{m_{da}} - \widetilde{m_{da}}^2}{c}\right) & \text{for humid air if } \alpha_{ice}=0\\
		\xi k_{par} + (1-\xi) k_{ser}  & \text{otherwise for frost}
	\end{cases}
	\label{Eq:thermal conductivity mixture}
\end{equation}
where, $\widetilde{m_{wv}}=\frac{Y \overline{M}}{M_{wv}}$ and $\widetilde{m_{da}}=\frac{(1-Y)\overline{M}}{M_{da}}$ are molar fractions of water vapor and dry air respectively and $c=2.75$. Average molar mass of mixture: $\overline{M}= \left[\frac{Y}{M_{wv}} + \frac{(1-Y)}{M_{da}}\right]^{-1}$. Parallel and serial values of conductivity are estimated as $k_{par}=\left(1-\frac{\rho_m}{\rho_{ice}}\right) k_{ha} + \frac{\rho_m}{\rho_{ice}} k_{ice}$ and $k_{ser}=\left[ \left(1-\frac{\rho_m}{\rho_{ice}}\right) \frac{1}{k_{ha}} + \frac{\rho_m}{\rho_{ice}} \frac{1}{k_{ice}} \right]^{-1}$ \cite{bartrons2019fixed,sanders1974influence,na2004new} and the parameter $\xi$ is given by:
\begin{equation*}
	\xi=
	\begin{cases}
		0.283 + \exp(-0.020 \rho_m) & \text{if } -10 < T_w < -4 ^\text{o} \text{C}\\
		0.140 + 0.919\exp(-0.0142 \rho_m) & \text{if } -21 < T_w < -10 ^\text{o} \text{C}\\
		0.0107 + 0.419\exp(-0.00424 \rho_m) & \text{if } T_w <-21 ^\text{o} \text{C and } \rho_m < 200 \text{ kg/m}^3\\
		0.005 \rho_m (0.0107 + 0.419\exp(-0.00424 \rho_m)) & \text{if } T_w <-21 ^\text{o} \text{C and } \rho_m > 200 \text{ kg/m}^3
	\end{cases}
\end{equation*}
The Studnikov relation \cite{studnikov1970viscosity} is also used to estimate viscosity of the humid air. Temperature dependent viscosities of dry air and water vapor are available in various references \cite{ccengel2008introduction, teske2005viscosity}. Properties which are independent of temperature or have minimal dependence are listed in \cref{Tab:Temperature Independent Properties}.

\begin{table}[H]
	\centering
	\caption{Properties with Negligible Temperature Dependence}
	\resizebox{\textwidth}{!}{%
		\begin{tabular}{|c|c|c|c|c|}
			\hline
			Property & Symbol & Value & Unit & Reference \\ \hline
			Density of Ice & $\rho_{ice}$ & 919 & kg/m$^3$ & \cite{harvey2016properties} \\ \hline
			Latent heat of sublimation for water & $L_{sub}$ & 2.841E6 & J/kg & \cite{stewart2009physical} \\ \hline
			Thermal conductivity of ice & $k_{ice}$ & 2.3 & W/m--K & \cite{harvey2016properties} \\ \hline
			Specific heat capacity of ice & $C_{p_{ice}}$ & 2000 & J/Kg--K & \cite{harvey2016properties} \\ \hline
			Specific heat capacity of air & $C_{p_{da}}$ & 1006 & J/Kg--K & \cite{lemmon2016properties} \\ \hline
			Thermal conductivity of water vapor & $k_{wv}$ & 0.016 & W/m--K & \cite{lide2005properties} \\ \hline
			Molecular mass of water & $M_{wv}$ & 18E--3 & Kg/mol & \\ \hline
			Molecular mass of dry air & $M_{da}$ & 29E--3 & Kg/mol & \\ \hline
		\end{tabular}%
	}
	\label{Tab:Temperature Independent Properties}
\end{table}

\subsection{Incorporation of Surface Wettability} \label{Sec:Nucleation Theory}
\begin{figure}[H]
	\centering
	\includegraphics[width=0.6\textwidth]{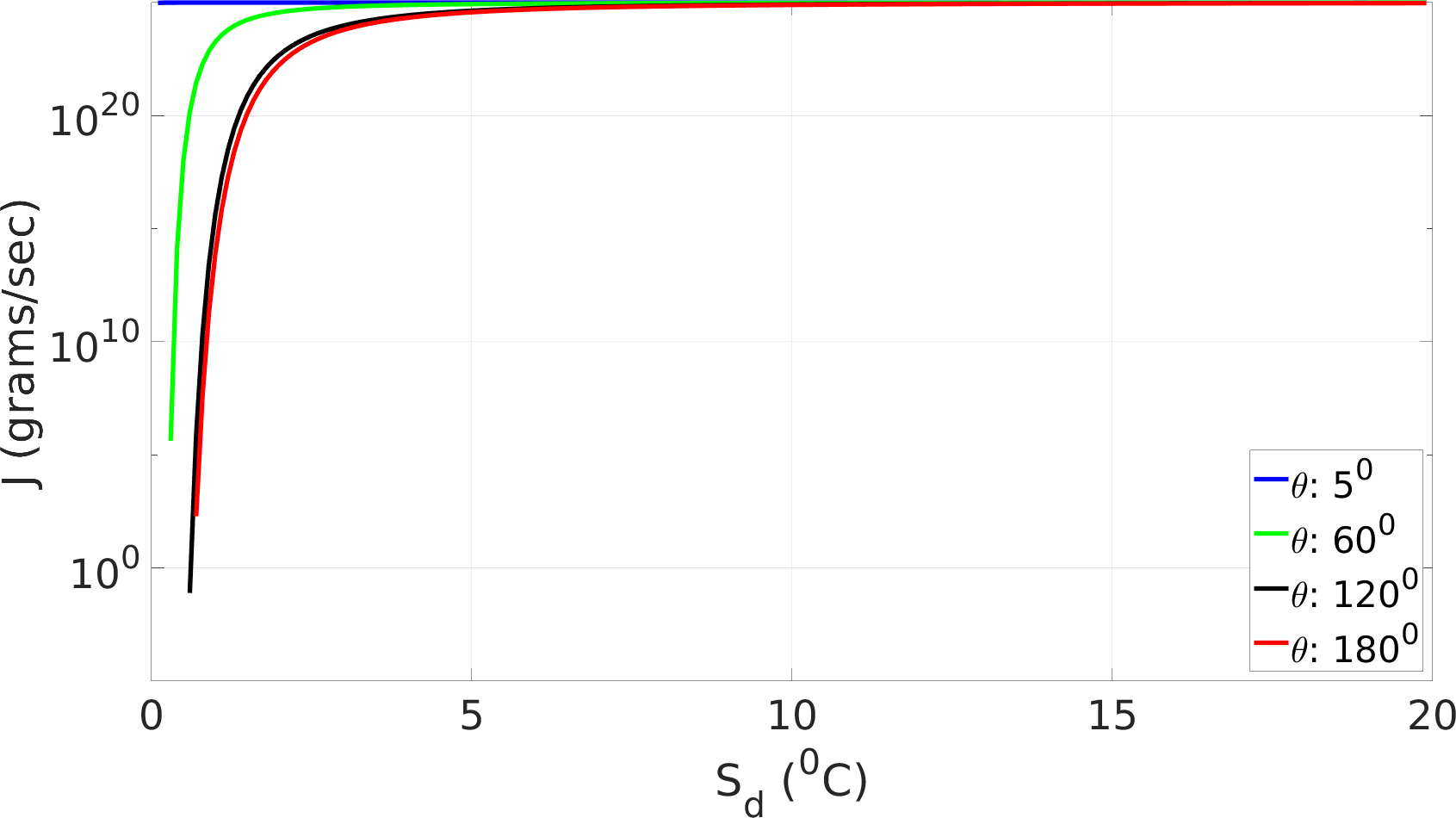}
	\caption{Nucleation Rate for Homogeneous and Heterogeneous Nucleation}
	\label{Fig:nucleation plots rate}
\end{figure}

Based on the nucleation theory, the heterogeneous nucleation rate $J_{het}$ can be modified from the homogeneous nucleation rate $J_{hom}$ by including the contact angle function $F(\theta)$, which converts the free energy change of the critical nucleus into a spherical cap shape induced by the contact angle $\theta$, as shown in \cref{Eq:nucleation 1,Eq:nucleation 2}, where, $v_m$ is the molecular volume, $\sigma$ is the surface tension and $k_b$ is the Boltzmann constant. The supersaturation $S$ is defined as the ratio of the vapor pressure ($p_{wv}$) and the equilibrium vapor pressure ($p_{ve}$): $S = {p_{wv}}/{p_{ve}}$. The kinetic prefactor ($K$) is a slight function of vapor pressure and temperature \cite{becker1935kinetische}, in most case it is assumed to be a constant of 1E21 $\sim$ 1E25 \cite{beysens2006dew, liu2000heterogeneous, iwamatsu2011heterogeneous}. In this work, $K$ is assumed to be 1E25 for the simulation.
\begin{equation}
	J_{het} = K \exp \left(F(\theta)\frac{-16 \pi v_m^2 \sigma^3}{3 (k_b T)^3 \ln(S)^2}\right)
	\label{Eq:nucleation 1}
\end{equation}
\begin{equation}
	F(\theta)= \left(\frac{1-\cos (\theta)}{2}\right)^2 (2+\cos (\theta))
	\label{Eq:nucleation 2}
\end{equation}
In \cref{Eq:nucleation 1}, the heterogeneous nucleation rate $J_{het}$ is affected by the temperature, supersaturation and surface wettability. The nucleation rate $J_{het}$ was plotted in \cref{Fig:nucleation plots rate}, with different supersaturation degrees Sd and contact angles at $20^\text{o}$C surrounding temperature, as well as assuming constant $K=10^{25}$. Instead of supersaturation $S = {p_{wv}}/{p_{ve}}$ in \cref{Eq:nucleation 1}, the supersaturation degree $S_d$ is the temperature difference at equilibrium vapor pressure pve and actual vapor pressure pv in saturation, $S_d = T_{sat}(p_{ve}) - T_{sat}(p_{wv})$. When the supersaturation degree increases, the nucleation rate also becomes larger, therefore nucleation process is prone to start.

As shown in \cref{Fig:nucleation plots rate}, for homogeneous nucleation ($\theta=180^\text{o}$), the nucleation process will not initiate in saturation vapor pressure ($S_d=0^\text{o}$C) as the nucleation rate is extremely small. In fact, the nucleation will start at $S_d=20^\text{o}$C \cite{gupta1946report} with considerable nucleation rate, which is also implied in \cref{Fig:nucleation plots rate}. For heterogeneous nucleation, a larger nucleation rate is achieved for surface with smaller contact angle. The heterogeneous nucleation rate $J_{het}$ decreases as $F(\theta)$ increases under the same temperature and supersaturation. Because $F(\theta)$ is positively correlated to the surface contact angle $\theta$, the nucleation rate $J_{het}$ is smaller for hydrophobic surface, compared with the hydrophilic surface. For example, ice surface is perfectly superhydrophilic \cite{knight1971experiments} (assuming contact angle $\theta=5^\text{o}$), which requires a lower supersaturation degree to start the nucleation.
\par With the critical nucleation rate $J_c$ assigned, the starting of nucleation process can be determined and the corresponding supersaturation can be calculated which reflects the surface wettability effect on the frosting process.

\subsection{Numerical Method}
In this section, we describe the numerical method with boundary conditions for solution of the governing equations discussed before. The conservation equations can be written as a scalar transport equation with unsteady, convection, diffusion and source terms:
\begin{equation}
	 \frac{\partial (\rho \phi)}{\partial t} + \nabla \bullet(\rho \bm{u} \phi) = \nabla \bullet ( \Gamma \nabla \phi) + S_{\phi}
	\label{Eq:scalar_transport}
\end{equation}
where, $\phi$ is any scalar field (velocity components, enthalpy etc.), $\Gamma$ is the diffusion coefficient, and $S_{\phi}$ is the source term. \Cref{Eq:scalar_transport} is integrated over a control volume to obtain \cref{Eq:scalar_transport_integrated}. The volume integral is converted to surface integral in the convection and diffusion terms using divergence theorem.
\begin{equation}
	\iiint_V \frac{\partial (\rho \phi)}{\partial t} dV + \oiint_S \rho \bm{u} \bullet \hat{n} \phi dS = \oiint_S \Gamma \nabla \phi \bullet \hat{n} dS + \iiint_V S_{\phi} dV
	\label{Eq:scalar_transport_integrated}
\end{equation}
Finite volume discretization approximates surface integral by summation over all the faces whereas, volume integrands are estimated as cell centered values multiplied by volume.
\begin{equation}
	\frac{\partial (\rho \phi)}{\partial t} \Delta V + \sum_{f} \left[\rho \bm{u} \bullet \hat{n} \phi \Delta A \right]_{f} = \sum_{f} \left[\Gamma \nabla \phi \bullet \hat{n} \Delta A \right]_{f} + S_{\phi} \Delta V
	\label{Eq:scalar_transport_integrated_FVM}
\end{equation}
where, $\Delta V$ is cell volume, $\Delta A$ is face area and $\hat{n}$ is outward facing normal of the face. These equations are discretized over unstructured grids so that complex geometries can be handled. Please refer to our previous work \cite{shahane2019finite, shahane2019numerical} for details of discretization of \cref{Eq:scalar_transport_integrated_FVM} over unstructured control volumes using the finite volume method.
\subsubsection{Pressure--Velocity Splitting} \label{Sec:Pressure--Velocity Coupling}
We extend the semi--implicit pressure splitting method \cite{harlow1965numerical} to solve coupled continuity and momentum equations (\cref{Eq:continuity,Eq:momentum x,Eq:momentum y}) using a semi--implicit approach. Since the advection term is non--linear, we iterate during each timestep. Let $r$ and $n$ denote iteration and timestep numbers respectively. The iterations are started ($r=0$) with the solution at timestep $n$.
\begin{equation}
	\frac{\rho_{ha}^n \alpha_{ha}^n \hat{u} \Delta V }{\Delta t} - \widehat{\mathcal{D}_u} + \Delta V K_d^n \hat{u} + \widehat{\mathcal{C}_u}= \frac{\rho_{ha}^n \alpha_{ha}^n  u^n \Delta V}{\Delta t} - \Delta V \left(\frac{\partial p}{\partial x}\right)^{r}
	\label{Eq:discrete momentum x iterate}
\end{equation}
where, $\widehat{\mathcal{D}_u} = \sum_{f} \left[\mu_{ha}^n \nabla \hat{u} \bullet \hat{n} \Delta A \right]_{f}$ is the diffusion term. The nonlinear convection term is linearized: $\widehat{\mathcal{C}_u} = \sum_{f} \left[(\rho^n \alpha_{ha}^n \bm{u}^r \bullet \hat{n} \Delta A) \hat{u}  \right]_{f}$. Term in the round bracket denotes mass flux which is estimated using velocity vector at the last iteration $r$. Similarly, the $Y-$momentum equation can be written as:
\begin{equation}
	\frac{\rho_{ha}^n \alpha_{ha}^n \hat{v} \Delta V }{\Delta t} - \widehat{\mathcal{D}_v} + \Delta V K_d^n \hat{v} + \widehat{\mathcal{C}_v}= \frac{\rho_{ha}^n \alpha_{ha}^n  v^n \Delta V}{\Delta t} - \Delta V \left(\frac{\partial p}{\partial y}\right)^{r}
	\label{Eq:discrete momentum y iterate}
\end{equation}
All the variables such as $\rho_{ha}$, $\alpha_{ha}$ and $K_d$ are estimated using the temperatures of the previous timestep ($n$). \Cref{Eq:discrete momentum x iterate,Eq:discrete momentum y iterate} are assembled in the form of sparse linear systems with $\hat{u}$ and $\hat{v}$ as the unknowns respectively. Similar discretization of the momentum \cref{Eq:momentum x} at new timestep $n+1$ gives:
\begin{equation}
	\frac{\rho_{ha}^n \alpha_{ha}^n u^{n+1} \Delta V }{\Delta t} - \mathcal{D}_u^{n+1} + K_d^n u^{n+1} \Delta V + \mathcal{C}_u^{n+1}= \frac{\rho_{ha}^n \alpha_{ha}^n u^n \Delta V}{\Delta t} - \left(\frac{\partial p}{\partial x}\right)^{n+1} \Delta V
	\label{Eq:discrete momentum x}
\end{equation}
Subtracting \cref{Eq:discrete momentum x iterate} from \cref{Eq:discrete momentum x} and neglecting the differences between terms at $r+1$ and $n+1$ of the diffusion, convection and Darcy drag terms:
\begin{equation}
	\frac{\rho_{ha}^n \alpha_{ha}^n (u^{r+1} - \hat{u})}{\Delta t} = -\frac{\partial (p^{r+1} - p^r)}{\partial x} = -\frac{\partial p'}{\partial x}
	\label{Eq:vel corr x}
\end{equation}
where, $p' = p^{r+1} - p^r$ is the correction in pressure. Note that $u^{n+1}$ is written as $u^{r+1}$ since it may not be the converged velocity. After the iterations converge, we set $u^{n+1} = u^{r+1}$. The correction equation for velocity in the $Y$ direction can be derived similarly by performing the previous steps on the $Y$ momentum \cref{Eq:momentum y}:
\begin{equation}
	\frac{\rho_{ha}^n \alpha_{ha}^n (v^{r+1} - \hat{v})}{\Delta t} = -\frac{\partial (p^{r+1} - p^r)}{\partial y} = -\frac{\partial p'}{\partial y}
	\label{Eq:vel corr y}
\end{equation}
Differentiating \cref{Eq:vel corr x} with $x$, \cref{Eq:vel corr y} with $y$ and adding the subsequent equations gives:
\begin{equation}
	\left[\frac{\partial (\rho_{ha}^n \alpha_{ha}^n u^{r+1}) }{\partial x} + \frac{\partial (\rho_{ha}^n \alpha_{ha}^n v^{r+1}) }{\partial y}\right] - \frac{\partial (\rho_{ha}^n \alpha_{ha}^n \hat{u})}{\partial x} - \frac{\partial (\rho_{ha}^n \alpha_{ha}^n \hat{v} )}{\partial y}
	= -\Delta t \nabla^2 p'
	\label{Eq:derivative vel corr add}
\end{equation}
Imposing the continuity \cref{Eq:continuity} on $u^{r+1}$ and $v^{r+1}$ gives:
\begin{equation}
	\left[\frac{\partial (\rho_{ha}^n \alpha_{ha}^n u^{r+1}) }{\partial x} + \frac{\partial (\rho_{ha}^n \alpha_{ha}^n v^{r+1}) }{\partial y}\right] = 0
	\label{Eq:discrete continuity}
\end{equation}
Substituting \cref{Eq:discrete continuity} into the \cref{Eq:derivative vel corr add} gives the pressure Poisson equation:
\begin{equation}
	\nabla^2 p' = \frac{1}{\Delta t}\left[\frac{\partial (\rho_{ha}^n \alpha_{ha}^n \hat{u})}{\partial x} + \frac{\partial (\rho_{ha}^n \alpha_{ha}^n \hat{v})}{\partial y} \right] = \frac{1}{\Delta t} \nabla \bullet \left[\rho_{ha}^n \alpha_{ha}^n \bm{\hat{u}} \right]
	\label{Eq:pressure poisson}
\end{equation}
\Cref{Eq:pressure poisson} is also integrated over the control volume and solved by finite volume method. Pressure correction equation with over--relaxation $\omega$ set to 1.4 is given by:
\begin{equation}
	p^{r+1} = p^r + \omega p'
	\label{Eq:pressure corr}
\end{equation}
The entire algorithm is described in \cref{Sec:Complete Solution Algorithm}.

\subsubsection{Equation for Conservation of Energy}
\Cref{Eq:enthalpy conservation} is expressed in terms of temperature as the unknown. The discretized in time as follows:
\begin{equation}
	\frac{\rho_m^n C_{p_m}^n T^{n+1} \Delta V }{\Delta t} - \mathcal{D}_T^{n+1} - \mathcal{H}_T^{n+1} + \mathcal{C}_T^{n+1} = \frac{\rho_m^n C_{p_m}^n T^n \Delta V}{\Delta t}
	\label{Eq:discrete enthalypy conservation}
\end{equation}
where, $\mathcal{D}_T$, $\mathcal{H}_T$ and $\mathcal{C}_T$ denote discrete diffusion of temperature, transport of enthalpy due to diffusion of water vapor in air and convection terms respectively. The convection term is linear in the unknown temperature ($T^{n+1}$) since the mass flux at $n+1$ is already estimated. The enthalpy transport term is linearized as follows:
\begin{equation}
	\mathcal{H}_T^{n+1} = \sum_{f} \left[(C_{p_{wv}}-C_{p_{da}})^n T^{n+1} \tau^n D_{wv}^n \rho_{ha}^n (\nabla Y)^n \bullet \hat{n} \Delta A \right]_{f}
	\label{Eq:discrete enthalypy transport}
\end{equation}
All terms in \cref{Eq:discrete enthalypy transport} except temperature are estimated from last timestep and thus, are added as coefficients in the discrete linear system.
\subsubsection{Equation for Conservation of Mass of Water}
\Cref{Eq:water conservation} models conservation of water which is solved to estimate the humidity mass fraction ($Y$) and volume fraction of ice ($\alpha_{ice}$). We discretize the \cref{Eq:water conservation} in terms of $Y$ and thus, an additional relation is needed to estimate the term with temporal derivative of ice volume fraction (${\partial (\rho_{ice} \alpha_{ice})}/{\partial t}$). Let $Y^n$ and $T^n$ denote humidity fraction and temperature at previous time step ($n$). Then energy \cref{Eq:enthalpy conservation} is solved to get the current time step estimate of temperature ($T^{n+1}$). Let $Y_{sat}^{n+1}$ denote saturation vapor content at $T^{n+1}$. $\alpha_{ice}$ has to be updated only if $Y^{n+1} > Y_{sat}^{n+1}$. Mass of water vapor in the  volume $\Delta V$ is $Y^{n+1} \rho_{ha}^n \Delta V (1-\alpha_{ice}^n)$. Thus, mass of water vapor which condenses is $(Y^{n+1}-Y_{sat}^{n+1}) \rho_{ha}^n \Delta V (1-\alpha_{ice}^n)$. Volume of this condensed ice is: ${(Y^{n+1}-Y_{sat}^{n+1}) \rho_{ha}^n \Delta V (1-\alpha_{ice}^n)}/{\rho_{ice}}$ and thus, the volume fraction is: ${(Y^{n+1}-Y_{sat}^{n+1}) \rho_{ha}^n (1-\alpha_{ice}^n)}/{\rho_{ice}}$. Hence, new value of volume fraction of ice ($\alpha_{ice}^{n+1}$) is given by:
\begin{equation}
	\alpha_{ice}^{n+1} =
	\begin{cases}
		\alpha_{ice}^n + \frac{(Y^{n+1}-Y_{sat}^{n+1}) \rho_{ha}^n (1-\alpha_{ice}^n)}{\rho_{ice}} & \text{if } Y^{n+1} > Y_{sat}^{n+1}\\
		\alpha_{ice}^n  & \text{otherwise}
	\end{cases}
	\label{Eq:ice volume fraction}
\end{equation}
In the region of saturation (if $Y^{n+1} > Y_{sat}^{n+1}$), substituting \cref{Eq:ice volume fraction} in the temporal derivative of ice volume fraction gives:
\begin{equation}
	\begin{aligned}
		\frac{\partial(\rho_{ice} \alpha)}{\partial t}&\approx \frac{\rho_{ice} (\alpha_{ice}^{n+1}- \alpha_{ice}^n)} {\Delta t}
		= \frac{(Y^{n+1}-Y_{sat}^{n+1}) \rho_{ha}^n (1-\alpha_{ice}^n)}{\Delta t}\\
		&= \frac{Y^{n+1} \rho_{ha}^n (1-\alpha_{ice}^n)}{\Delta t} - \frac{Y_{sat}^{n+1} \rho_{ha}^n (1-\alpha_{ice}^n)}{\Delta t}
	\end{aligned}
	\label{Eq:ice volume fraction time derivative}
\end{equation}
First term is modeled implicitly with ${\rho_{ha}^n (1-\alpha_{ice}^n)}/{\Delta t}$ as coefficient of the diagonal term in the discretized \cref{Eq:water conservation}. The second term is taken as an explicit source term. If $Y^{n+1} \leq Y_{sat}^{n+1}$, ${\partial(\rho_i \alpha)}/{\partial t}=0$. Finally, the discretized water conservation \cref{Eq:water conservation} takes the following form:
\begin{equation}
	Y^{n+1} \left(\frac{\rho_{ha}^n + \rho_{ha}^n (1-\alpha_{ice}^n) \Delta V}{\Delta t}\right) - \mathcal{D}_Y^{n+1} + \mathcal{C}_Y^{n+1}  = \frac{\rho_{ha}^n Y^n \Delta V }{\Delta t} + \frac{Y_{sat}^{n+1} \rho_{ha}^n (1-\alpha_{ice}^n) \Delta V}{\Delta t}
	\label{Eq:discrete water conservation}
\end{equation}
where, $\mathcal{D}_Y$ and $\mathcal{C}_Y$ denoting discrete diffusion and convection terms respectively are treated implicitly. The convection term is linear in the unknown vapor fraction ($Y^{n+1}$) since the mass flux at $n+1$ is already estimated.
\par Since the energy \cref{Eq:enthalpy conservation} is solved for $T^{n+1}$ before solving the water conservation \cref{Eq:water conservation}, $Y_{sat}^{n+1}$ is estimated as a function of $T^{n+1}$. Supersaturation temperature is estimated using the model discussed in \cref{Sec:Nucleation Theory}. First, we solve \cref{Eq:nucleation 2} for $S$ assuming $J_{het} = K/2$ by substituting the value of local temperature and given surface contact angle. Then the vapor pressure ($p_{wv}$) is computed using the relation $S = {p_{wv}}/{p_{ve}}$ which in turn gives the temperature at saturation pressure of $p_{wv}$ by solving the following equation:
\begin{equation}
	p(T) = 610.94 \exp \left(\frac{17.625 T}{T + 243.04}\right)
	\label{Eq:nucleation 4}
\end{equation}
where, temperature $T$ is measured in Kelvin and pressure $p(T)$ is in Pascal.
The difference between this temperature and actual local temperature ($T^{n+1}$) gives the value of supersaturation ($\Delta T^{ss}$). Note that the supersaturation temperature is nonzero only for the control volume adjacent to the cold surface. In the interior of the domain, its value is set to zero since the effect of contact angle is not observed away from the cold surface. $Y_{sat}^{n+1}$ is then computed using the vapor pressure ($p_{sat}$) at $T=T^{n+1}+\Delta T^{ss}$ \cite{handbook2001hvac}:
\begin{equation}
	Y_{da} = \frac{0.62198 p_{sat}}{p_{atm} - p_{sat}} \hspace{0.5cm} \text{thus,} \hspace{0.5cm}
	Y_{sat}^{n+1} = \frac{Y_{da}}{1+Y_{da}}
	\label{Eq:nucleation 5}
\end{equation}
where, $p_{atm}$ is the atmospheric pressure and $Y_{da}$ denotes mass of water vapor per unit mass of dry air.

\subsubsection{Implementation of Boundary Conditions}\label{Sec:Implementation of Boundary Conditions}
The following conditions are imposed at various domain boundaries for the primary variables ($u$, $v$, $p$, $T$ and $Y$) \cite{lee2003prediction, cui2011new_ate, kim2015frosting}:
\begin{itemize}
	\item \textbf{Inlet}: $u$, $v$, $T$ and $Y$ prescribed; ${\partial^2 p}/{\partial \hat{n}^2}=0$ where, $\hat{n}$ denotes the unit normal and ${\partial }/{\partial \hat{n}}$ stands for $\nabla  \bullet \hat{n}$.
	\item \textbf{Cold wall}: $u=0$, $v=0$, $T$ prescribed, ${\partial Y}/{\partial \hat{n}}=0$, ${\partial p}/{\partial \hat{n}}=0$
	\item \textbf{Insulated wall}: $u=0$, $v=0$, ${\partial T}/{\partial \hat{n}}=0$, ${\partial Y}/{\partial \hat{n}}=0$, ${\partial p}/{\partial \hat{n}}=0$
	\item \textbf{Outlet}: ${\partial u}/{\partial \hat{n}}=0$, ${\partial v}/{\partial \hat{n}}=0$, ${\partial T}/{\partial \hat{n}}=0$, ${\partial Y}/{\partial \hat{n}}=0$, ${\partial^2 p}/{\partial \hat{n}^2}=0$
\end{itemize}
Note that the pressure boundary conditions at the inlet and outlet are difficult to estimate. Hence, we assume the conditions of a fully developed flow. \Cref{Eq:discrete momentum x iterate,Eq:discrete momentum y iterate} are first solved by imposing Dirichlet boundary condition at outlet with velocities at last iteration ($u^r$ and $v^r$). Velocities in the domain interior are corrected to estimate $u^{n+1}$ and $v^{n+1}$ using the pressure gradients (\cref{Eq:vel corr x,Eq:vel corr y}). Then the velocities ($u^{n+1}$ and $v^{n+1}$) at the inlet and wall boundaries are set as described above. At the outlet boundary, velocities are estimated by solving ${\partial u^{n+1}}/{\partial \hat{n}}=0$ and ${\partial v^{n+1}}/{\partial \hat{n}}=0$. After imposing these conditions, the outlet velocities are scaled using the flow rate at inlet in order to ensure global flux balance.

\subsubsection{Complete Solution Algorithm} \label{Sec:Complete Solution Algorithm}
In this section, we list down all the steps of the entire algorithm. Given the values of the primary variables ($u$, $v$, $p$, $T$ and $Y$) at timestep $n$, the following steps are performed to estimate their values at the next timestep $n+1$:
\begin{enumerate}
	\item Set values of density ($\rho_{da}^n$, $\rho_{ha}^n$, $\rho_m^n$), thermal conductivity ($k_m^n$), specific heat capacity ($C_{p_{da}}^n$, $C_{p_{wv}}^n$), viscosity ($\mu_{ha}^n$), diffusivity of water vapor in air ($D_{wv}^n$), diffusion resistance factor ($\tau^n$) and Darcy drag coefficient ($K_d^n$) using \cref{Eq:Darcy K,Eq:humid air density,Eq:mean density,Eq:thermal conductivity mixture,Eq:diffusion resistance,Eq:specific heat capacities}.
	\item Initialize values of pressure and velocities at $r=0$ using last timestep values at $n$ and iterate from $r$ to $r+1$ to estimate pressure and velocities at $n+1$:
	\begin{enumerate}
		\item Solve \cref{Eq:discrete momentum x iterate,Eq:discrete momentum y iterate} for $\hat{u}$ and $\hat{v}$ by imposing Dirichlet boundary conditions with values at last iteration ($r$). Central differencing and first order upwinding is used for discretization of the diffusion and convection terms respectively.
		\item Calculate error in non--dimensional mass flux $\left( \sum_{f} \left[ \rho_{ha}^n \alpha_{ha}^n \bm{\hat{u}}  \bullet \hat{n} \Delta A \right]_{f} / \dot{m}_{in} \right)$ where, $\dot{m}_{in}$ denotes mass flux at inlet.
		\item Estimate pressure correction by solving the pressure Poisson \cref{Eq:pressure poisson} with appropriate boundary conditions described in \cref{Sec:Implementation of Boundary Conditions}. Correct pressure using \cref{Eq:pressure corr}.
		\item Estimate mass fluxes at the faces of the control volumes by applying the velocity correction \cref{Eq:vel corr x,Eq:vel corr y} at face centers
		\item Correct the cell centered velocities to $u^{r+1}$ and $v^{r+1}$ from the \cref{Eq:vel corr x,Eq:vel corr y} in the domain interior.
		\item Correct $u^{r+1}$ and $v^{r+1}$ at the boundaries as described in \cref{Sec:Implementation of Boundary Conditions}.
		\item If the error in mass flux computed in step 2(b) is less than tolerance (0.001 in this case), stop iterations, set $\bm{u^{n+1}}=\bm{u^{r+1}}$ and go to step 3. Otherwise, go back to step 2(a).
	\end{enumerate}
	\item Solve the energy conservation \cref{Eq:discrete enthalypy conservation} with \cref{Eq:discrete enthalypy transport} and estimate $T^{n+1}$.
	\item Solve the \cref{Eq:discrete water conservation} which conserves mass of water together with \cref{Eq:ice volume fraction,Eq:ice volume fraction time derivative} and obtain the values of $Y^{n+1}$ and $\alpha_{ice}^{n+1}$. Note that $\alpha_{ha}^{n+1}=1-\alpha_{ice}^{n+1}$. The supersaturation temperature is estimated here using the model described in \cref{Sec:Nucleation Theory}.
\end{enumerate}
The above steps are performed at each timestep. First order Euler method is used for time marching. The semi--implicit formulation is stable beyond a Courant number of unity which is defined as $Co = u_c \Delta t /\Delta x$. The inlet velocity is set as characteristic speed ($u_c$). We find that after first few timesteps, the velocity--pressure splitting approach (steps 2(a)--(g)) converges in a single iteration per timestep for Courant number of less than 10.
\section{Validation}
\begin{figure}[H]
	\centering
	\includegraphics[width=0.7\textwidth]{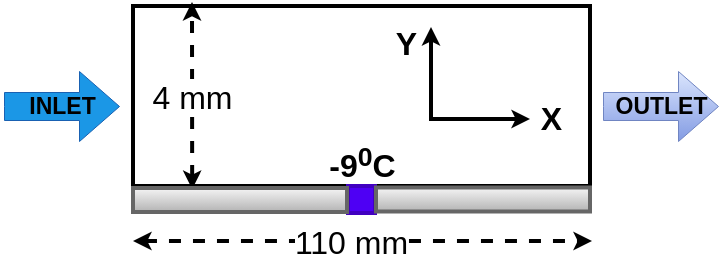}
	\caption{Schematic of Domain}
	\label{Fig:validation domain}
\end{figure}
We use the experimental results of \citet{kwon2006experimental} to validate this model. Two flat plates of dimension 110 mm by 100 mm are placed with 4 mm gap in between. A cooling source is placed at the center of the bottom plate to maintain a temperature close to $-9^\text{o}$C. Bottom plate temperatures and frost thickness are measured by \citet{kwon2006experimental} at various locations along the flow direction. We model this problem by a two dimensional domain with height 4 mm and length 110 mm (\cref{Fig:validation domain}). The documented temperatures in the paper along the bottom plate are imposed as boundary condition in this model. Inlet and outlet boundary conditions are set at the left and right ends of the domain respectively. An insulated wall boundary condition is applied at the top wall. The inlet velocity, temperature and humidity are set to 1.5 m/s, $-2^\text{o}$C and 3.65 $\text{g}_{\text{wv}}/\text{kg}_{\text{da}}$ respectively.

\begin{figure}[H]
	\centering
	\includegraphics[width=0.8\textwidth]{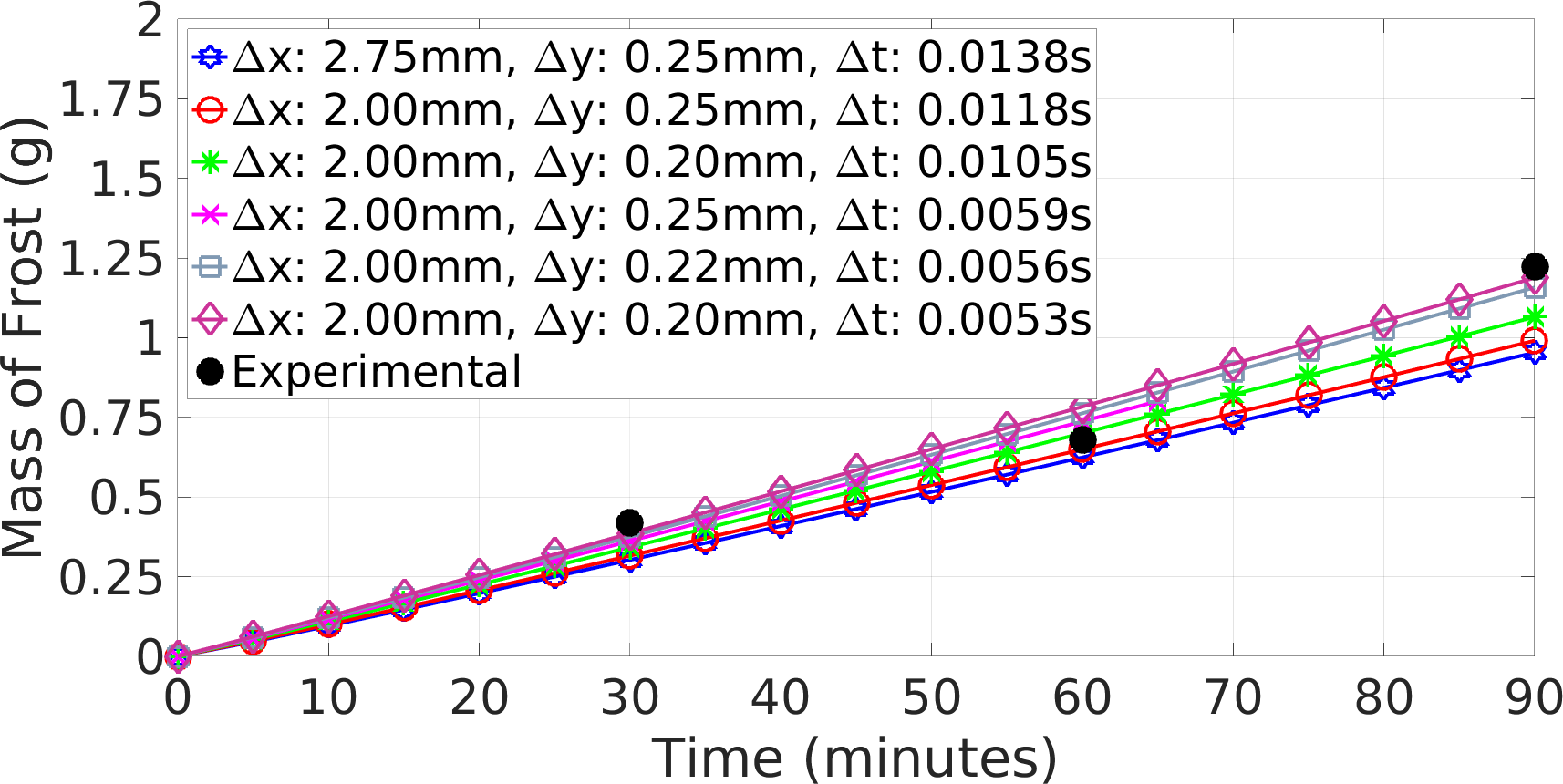}
	\caption{Grid Independence: Temporal Evolution of Frost Mass (g)}
	\label{Fig:validation Grid Independence}
\end{figure}
First step is to find a grid and timestep independent solution. Since the gradients along Y direction are strong, $\Delta y$ is kept smaller than $\Delta x$. \Cref{Fig:validation Grid Independence} plots mass of deposited ice with time for varying grid spacing and timestep. The experimental values documented by \citet{kwon2006experimental} at 30, 60 and 90 minutes are also plotted. It can be seen that the finer grid and timestep predictions overlap with the experiential data. The remaining results in this section are plotted using the finest grid with $\Delta x=2$ mm, $\Delta y=0.2$ mm and $\Delta t=0.0053$ s.
\par
\begin{figure}[H]
	\centering
	\begin{subfigure}[t]{0.49\textwidth}
		\includegraphics[width=\textwidth]{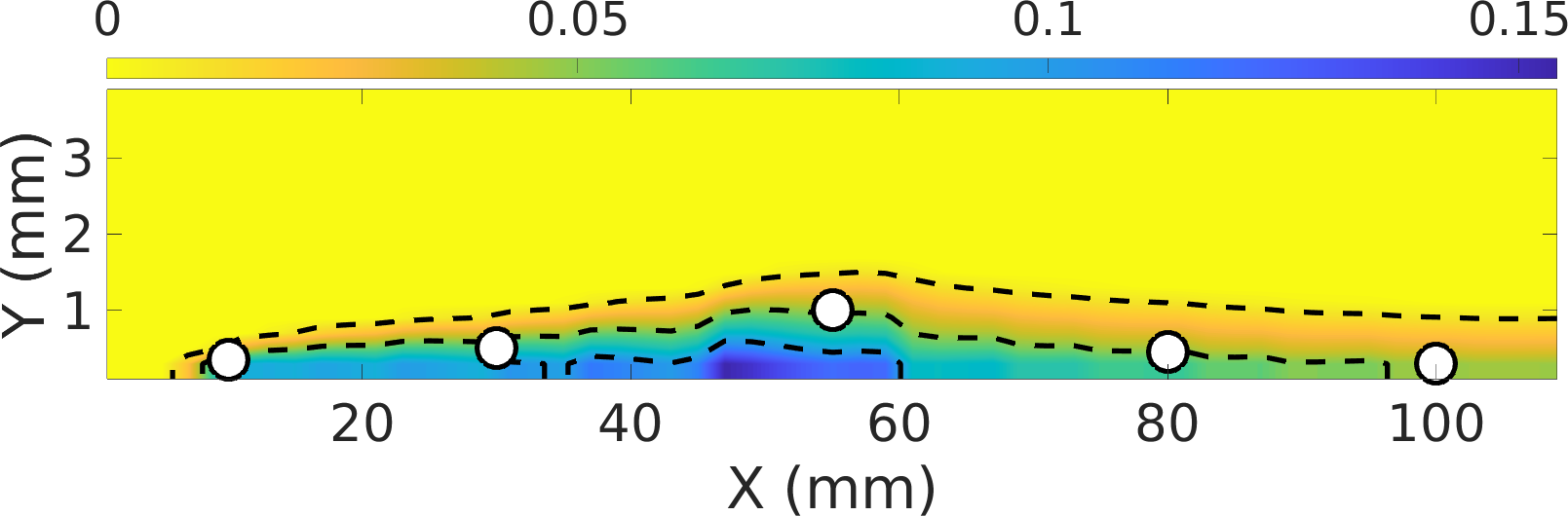}
		\caption{30 minutes}
	\end{subfigure}
	\begin{subfigure}[t]{0.49\textwidth}
		\includegraphics[width=\textwidth]{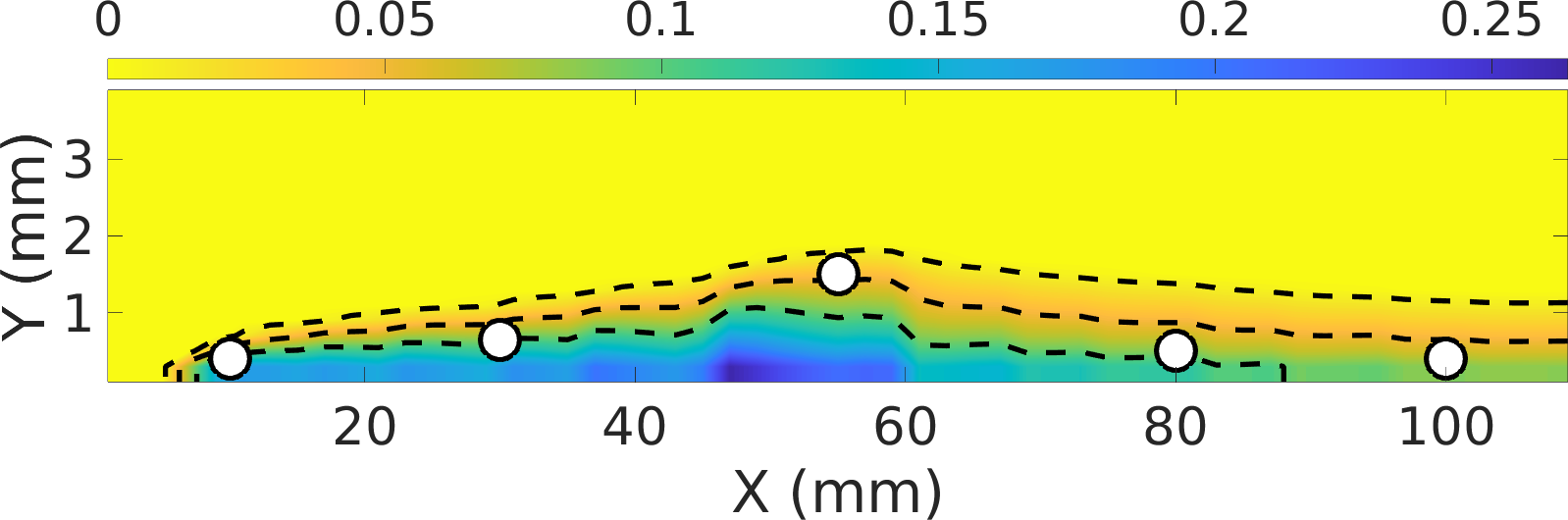}
		\caption{60 minutes}
	\end{subfigure}
	\begin{subfigure}[t]{0.49\textwidth}
		\includegraphics[width=\textwidth]{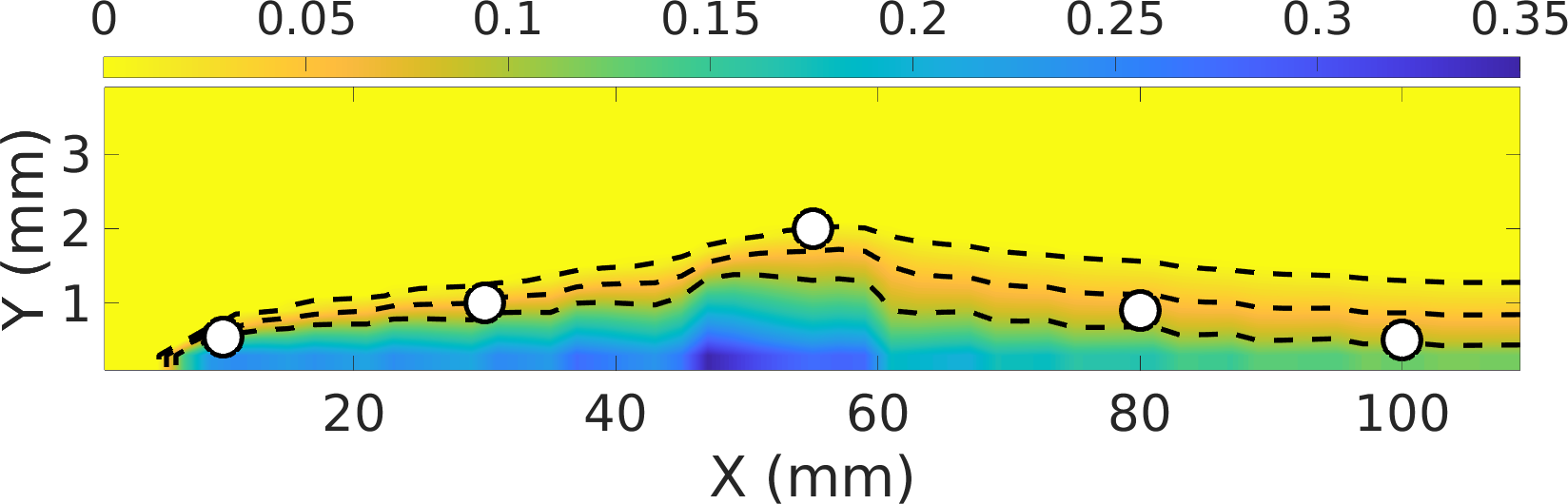}
		\caption{90 minutes}
	\end{subfigure}
	\caption{Volume Fraction of Ice (Markers: Experimental Frost Thickness \cite{kwon2006experimental}, Dashed Iso--lines of $\alpha_{ice}=$ 0.1, 0.05, 0.01)}
	\label{Fig:validation alpha}
\end{figure}

\begin{figure}[H]
	\centering
	\begin{subfigure}[t]{0.49\textwidth}
		\includegraphics[width=\textwidth]{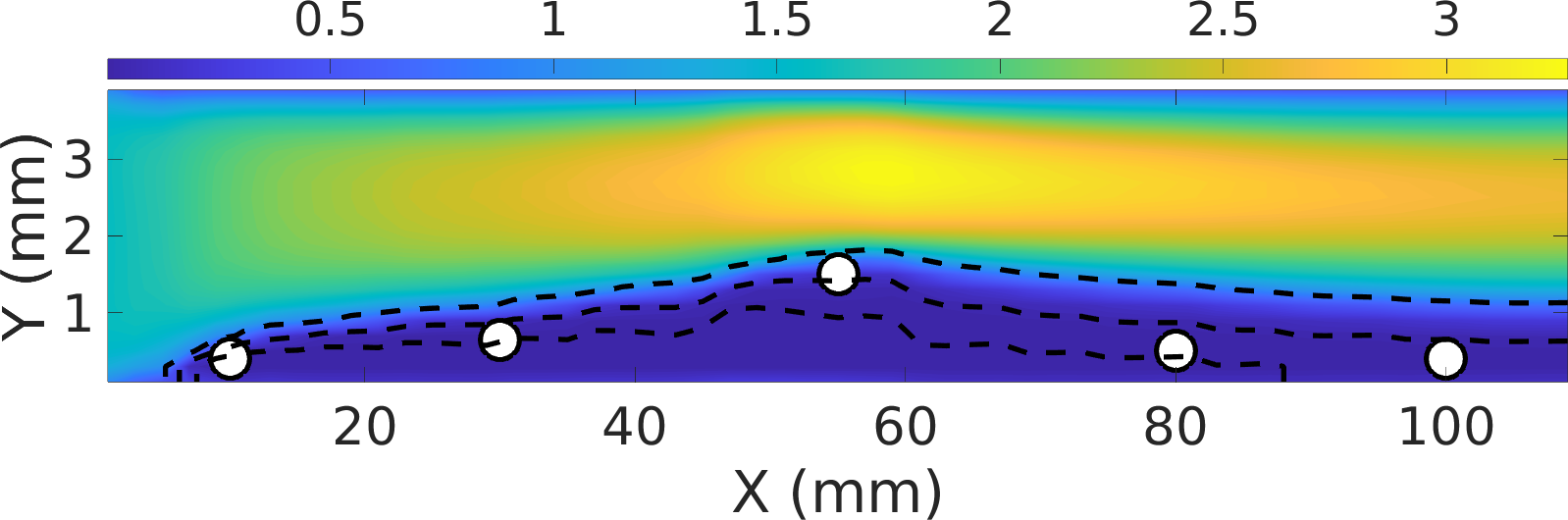}
		\caption{X Velocity (m/s)}
	\end{subfigure}
	\begin{subfigure}[t]{0.49\textwidth}
		\includegraphics[width=\textwidth]{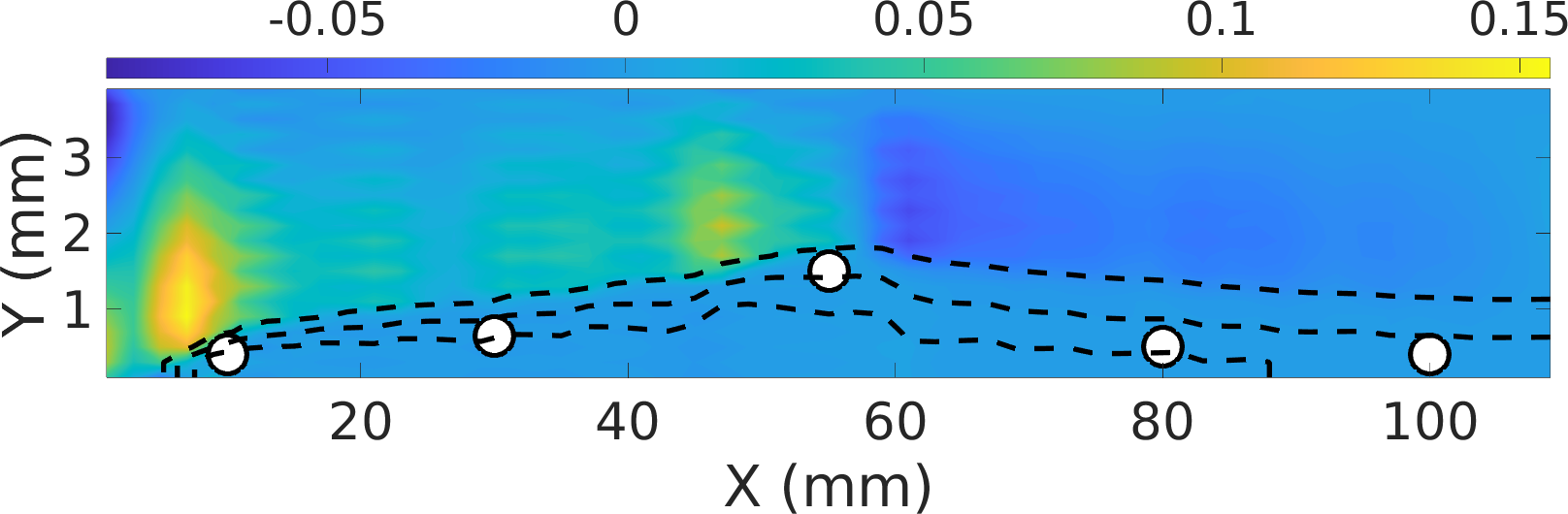}
		\caption{Y Velocity (m/s)}
	\end{subfigure}
	\begin{subfigure}[t]{0.49\textwidth}
		\includegraphics[width=\textwidth]{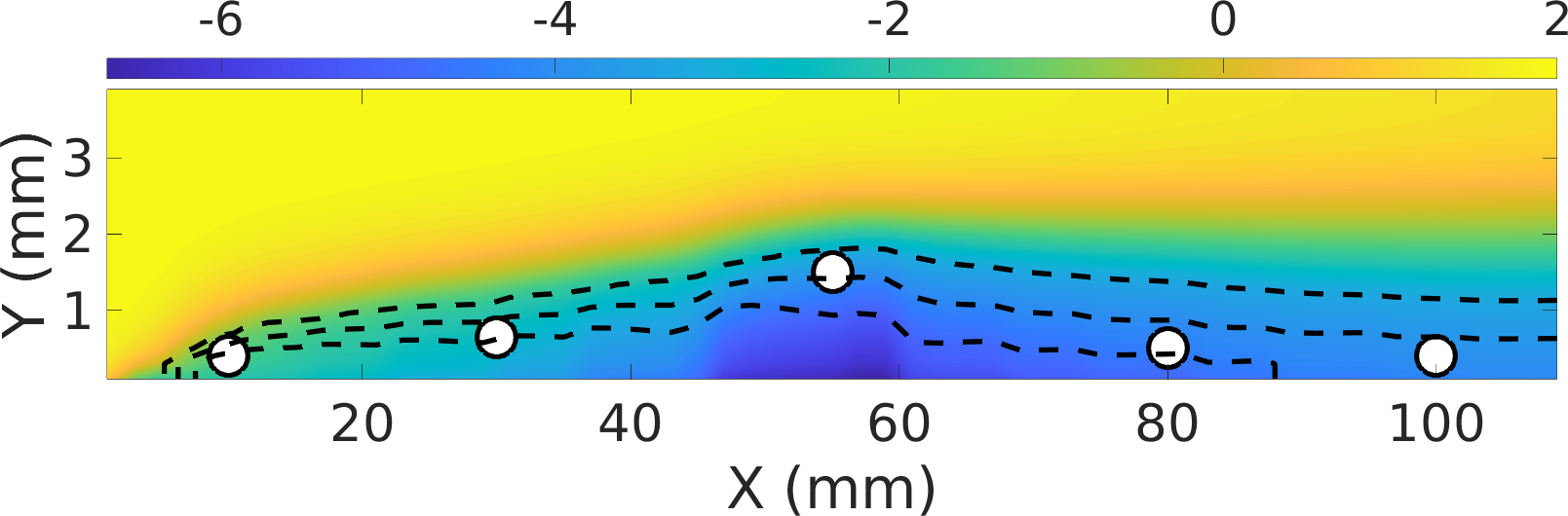}
		\caption{Temperature ($^\text{o}$C)}
	\end{subfigure}
	\begin{subfigure}[t]{0.49\textwidth}
		\includegraphics[width=\textwidth]{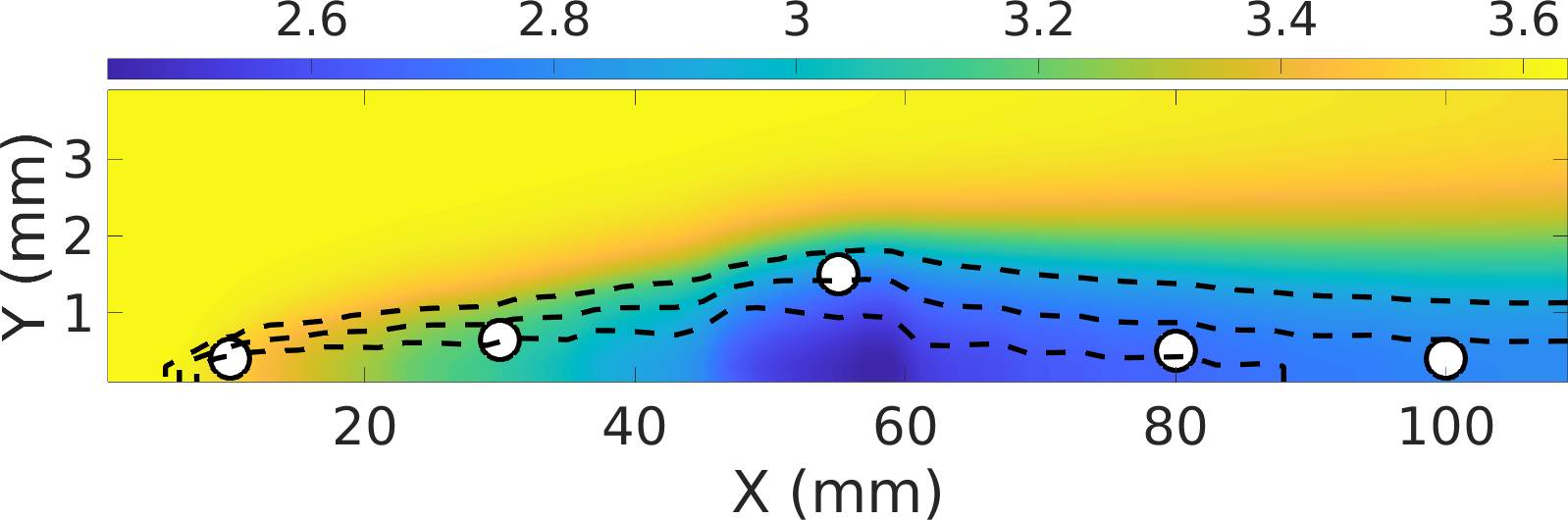}
		\caption{$Y$: Ratio of Water Vapor Mass to Humid Air Mass ($\text{g}_{\text{wv}}/\text{kg}_{\text{ha}}$)}
	\end{subfigure}
	\caption{60 minutes (Markers: Experimental Frost Thickness \cite{kwon2006experimental}, Dashed Iso--lines of $\alpha_{ice}=$ 0.1, 0.05, 0.01)}
	\label{Fig:validation 60 min}
\end{figure}

 \Cref{Fig:validation alpha} plots the contours of ice volume fraction ($\alpha_{ice}$) for 30, 60 and 90 minutes. Since the central region of the plate is cooled down, we see an increase followed by a decrease in frost thickness from inlet to outlet. The mixture model does not explicitly track the interface. Hence, we plot 3 dashed iso--lines corresponding to simulated values of $\alpha_{ice}=$ [0.1, 0.05, 0.01]. Experimental data of frost thickness logged by \citet{kwon2006experimental} is plotted as circular markers together with the contour plots. It can be seen that the experimental markers lie well within the simulated iso--lines.
Contours of velocity components, temperature and humidity fraction at 60 minutes are plotted in \cref{Fig:validation 60 min}. The experimental thickness and simulated iso--lines are again added to these plots for analysis. Velocities in the frost region are lower since it offers resistance to flow of humid air. Thus, it can be seen that the flow turns in the direction away from the cold plate due to higher frost deposition in the central region. Near the domain exit region, the flow again bends towards the plate since the frost thickness is lower. Moreover, due to deceleration of the flow in the frost region, the flow accelerates in the region of humid air in the domain center. The effect of cooling is observed in the contours of temperature and humidity fraction as well. Since the capacity of air to hold water vapor decreases with drop in temperature, the contours of humidity ratio are similar to temperature. In order to save space, we do not plot similar contours at 30 and 90 minutes.

\section{Results and Discussions}
\begin{figure}[H]
	\centering
	\begin{subfigure}[t]{0.5\textwidth}
		\includegraphics[width=\textwidth]{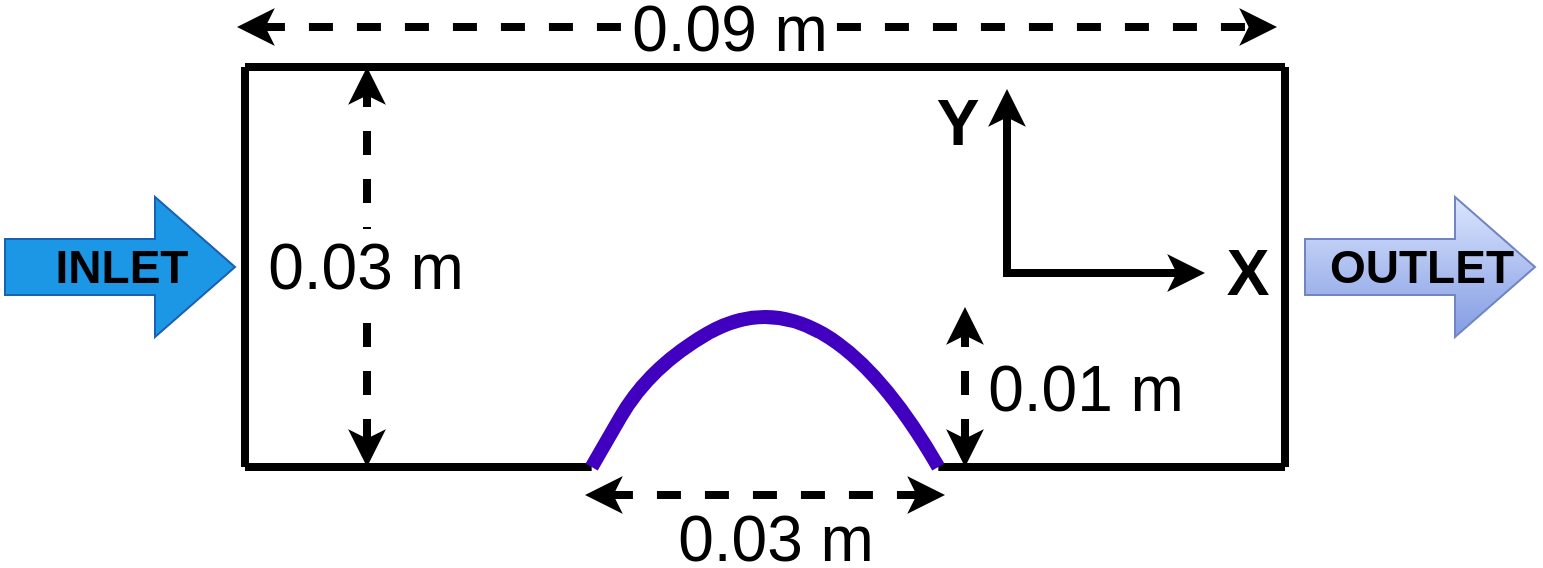}
		\caption{Geometry}
		\label{Fig:Schematic Sine Geometry}
	\end{subfigure}
	\hspace{0.05\textwidth}
	\begin{subfigure}[t]{0.43\textwidth}
		\includegraphics[width=\textwidth]{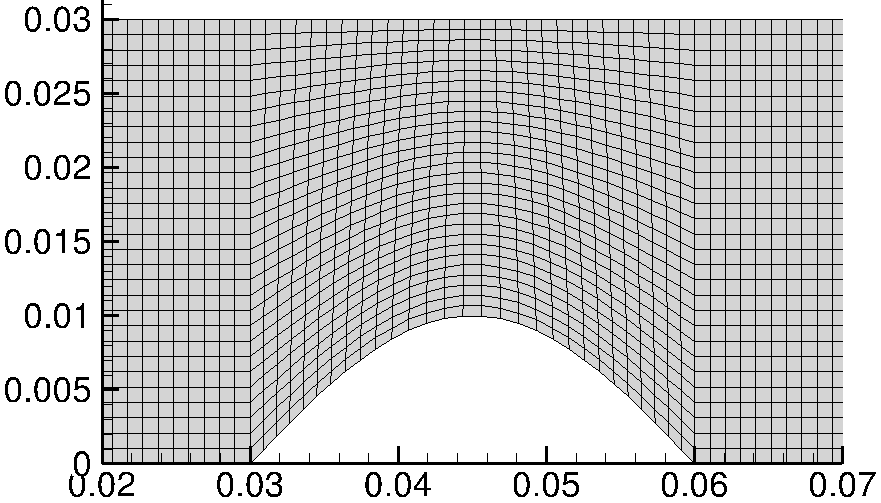}
		\caption{Grid: $\Delta x=0.85$ mm, 2800 elements}
		\label{Fig:Schematic Sine Grid}
	\end{subfigure}
	\caption{Schematic}
	\label{Fig:Schematic Sine}
\end{figure}

\begin{figure}[H]
	\centering
	\includegraphics[width=0.8\textwidth]{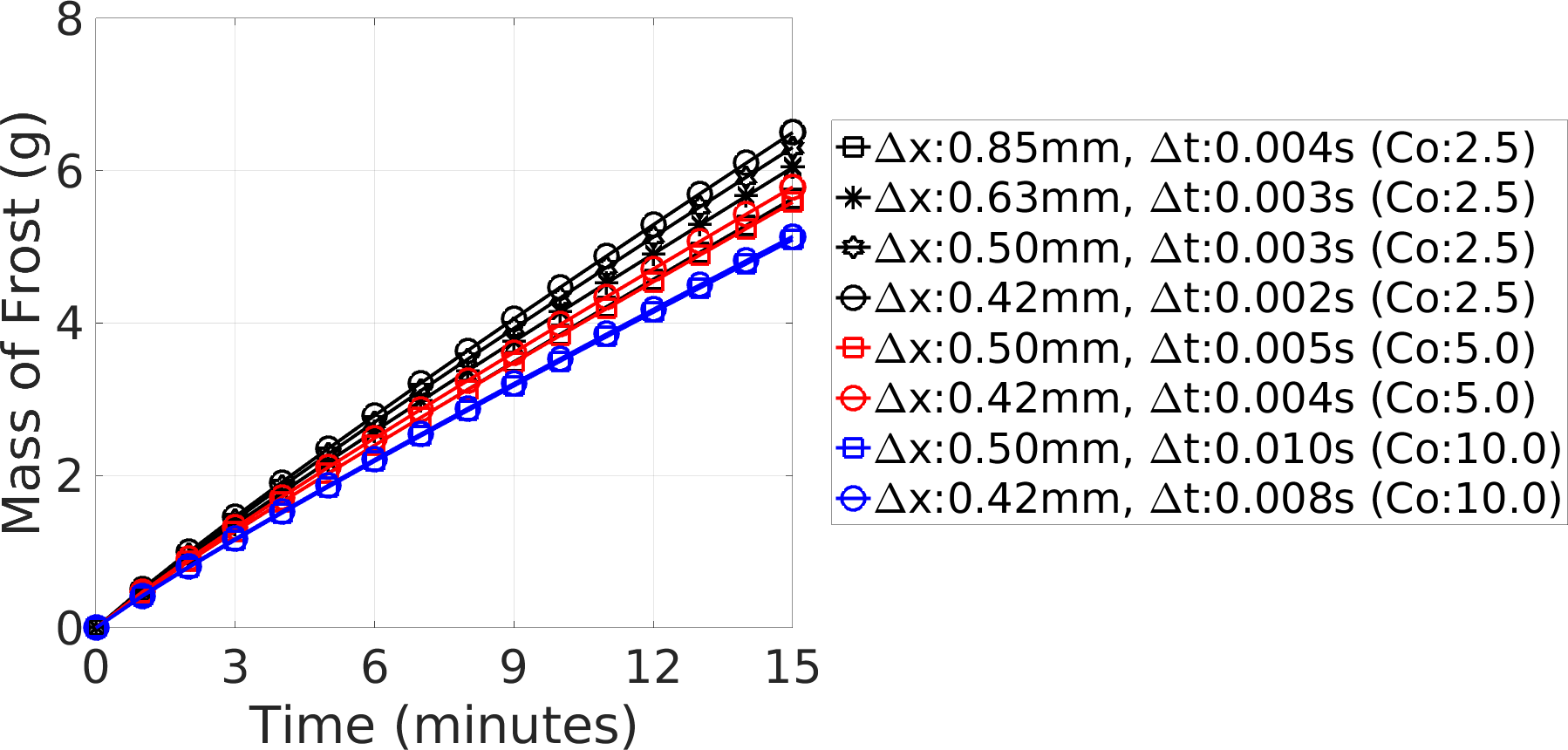}
	\caption{Grid and Timestep Independence: Temporal Evolution of Frost Mass (g)}
	\label{Fig:Grid Independence}
\end{figure}

In order to show the application of the numerical method introduced in this paper, we study frost growth on a two dimensional sinusoidal surface. \Cref{Fig:Schematic Sine Geometry} shows the domain geometry with dimensions. A half sine wave of amplitude 10 mm and width 30 mm is cooled to a sub--zero temperature and humid air is flown over it. \Cref{Fig:Schematic Sine Grid} shows a typical grid around the sine wave generated using GMSH \cite{geuzaine2009gmsh}. In the upstream and downstream region, a structured Cartesian grid is used. For this unstructured grid, $\Delta x$ is defined as the square root of the minimum area among all the control volumes.
The first step is to identify a reasonable grid size ($\Delta x$) and timestep value ($\Delta t$). For this analysis, we simulate only for 15 minutes. The semi--implicit formulation discussed before allows for convection Courant numbers above unity. The temporal evolution of mass of deposited frost is plotted in \cref{Fig:Grid Independence} for four grid resolutions and three convection Courant numbers. Asymptotic convergence is observed with grid and timestep refinement. We choose the grid spacing of $\Delta x=0.5$ mm and a Courant number of 2.5 for all further calculations.
Flow velocity, relative humidity and temperature of the humid air at the inlet as well as surface temperature are the four parameters which affect the rate of frost growth. In order to study the effect of these parameters, we define a baseline simulation and vary one parameter at a time keeping others the same. \Cref{tab:List of Simulations} lists down 9 simulations with B0 as the baseline. UI, HI, TI and TS stand for inlet velocity, humidity, temperature and surface temperature respectively. For the case UI, we perform 2 simulations: with velocity lower and higher then B0 marked with red color in the table. Other parameters between UI1 and UI2 are the same as B0. Similarly, other cases are defined. Frost growth is simulated for 2 hours.
\begin{table}[H]
	\centering
	\resizebox{\textwidth}{!}{%
	\begin{tabular}{|c|c|c|c|c|c||c|}
		\hline
		\begin{tabular}[c]{@{}c@{}}Sim. \\ Name\end{tabular} &
		\begin{tabular}[c]{@{}c@{}}Inlet \\ Vel. (m/s)\end{tabular} &
		\begin{tabular}[c]{@{}c@{}}Inlet Relative \\ Humidity (\%)\end{tabular} &
		\begin{tabular}[c]{@{}c@{}}Inlet \\ Temp. ($^\text{o}$C)\end{tabular} &
		\begin{tabular}[c]{@{}c@{}}Surface \\ Temp. ($^\text{o}$C)\end{tabular} &
		\begin{tabular}[c]{@{}c@{}}Contact \\ Angle (deg)\end{tabular} &
		\begin{tabular}[c]{@{}c@{}}Result: Frost \\ Rate (g/min)\end{tabular} \\ \hline
		B0  & 0.3 & 60 & 10 & -10 & 90 & 0.1043 \\ \hline
		UI1 & {\color{red}\textbf{0.1}} & 60 & 10 & -10 & 90 & 0.0629 \\ 
		UI2 & {\color{red}\textbf{0.5}} & 60 & 10 & -10 & 90 & 0.1294 \\ \hline
		HI1 & 0.3 & {\color{red}\textbf{40}} & 10 & -10 & 90 & 0.0322 \\ 
		HI2 & 0.3 & {\color{red}\textbf{80}} & 10 & -10 & 90 & 0.1684 \\ \hline
		TI1 & 0.3 & 60 & {\color{red}\textbf{5}}  & -10 & 90 & 0.0529 \\ 
		TI2 & 0.3 & 60 & {\color{red}\textbf{15}} & -10 & 90 & 0.1729 \\ \hline
		TS1 & 0.3 & 60 & 10 & {\color{red}\textbf{-15}} & 90 & 0.1312 \\ 
		TS2 & 0.3 & 60 & 10 & {\color{red}\textbf{-5}}  & 90 & 0.0628 \\ \hline
		CA1 & 0.3 & 60 & 10 & -10 & {\color{red}\textbf{30}} & 0.1062 \\ 
		CA2 & 0.3 & 60 & 10 & -10 & {\color{red}\textbf{150}}& 0.1051 \\ \hline
	\end{tabular}
	}
	\caption{List of Simulations}
	\label{tab:List of Simulations}
\end{table}

\begin{figure}[H]
	\centering
	\includegraphics[width=0.8\textwidth]{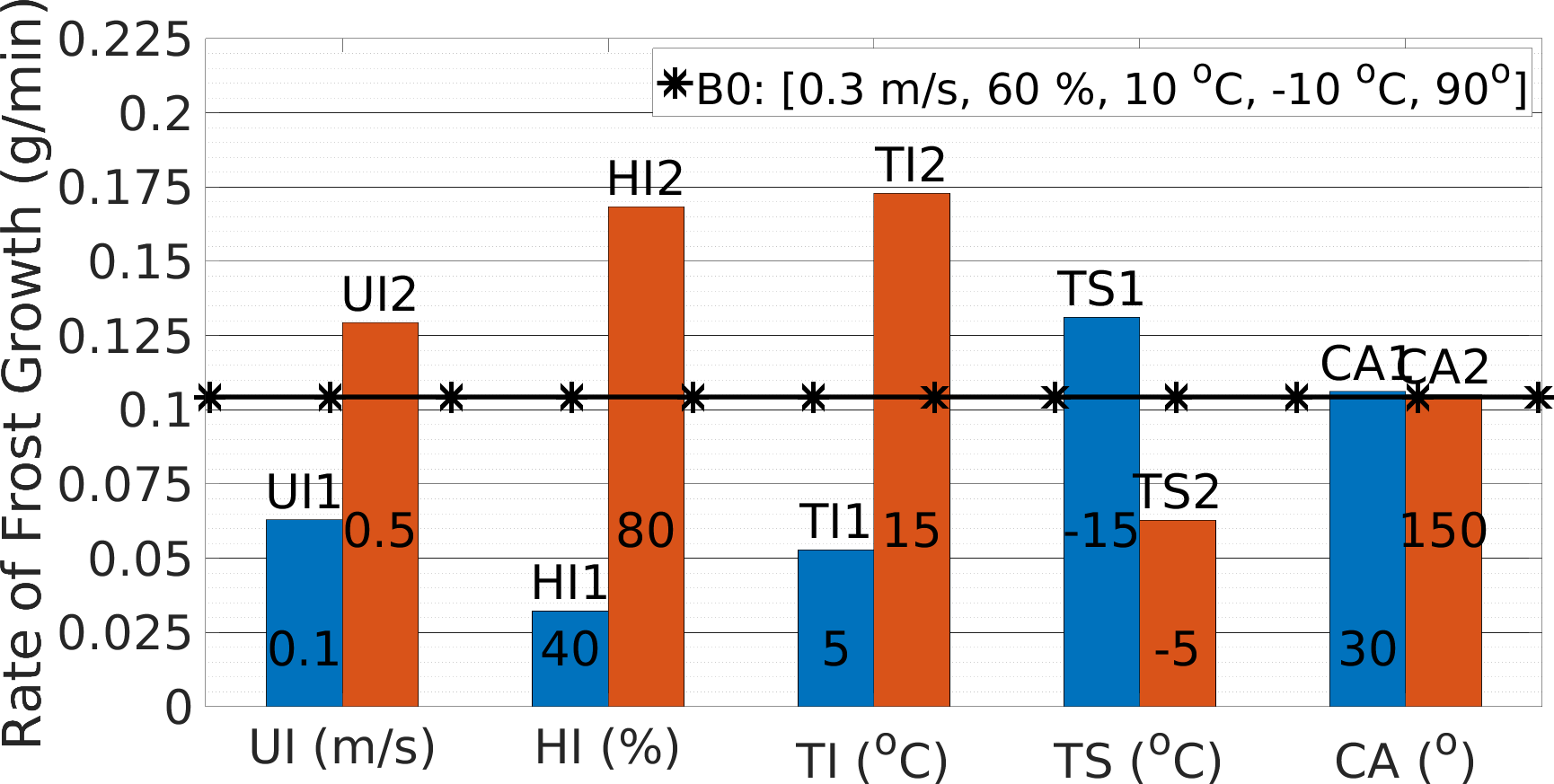}
	\caption{Rate of Growth of Frost Mass (g/min)}
	\label{Fig:Rate of Growth of Frost Mass}
\end{figure}
\Cref{Fig:Rate of Growth of Frost Mass} summarizes the effect of all the four parameters compared with the baseline case. We observe that the temporal evolution of mass of frost is almost linear (\cref{Fig:Grid Independence,Fig:validation Grid Independence}). Hence, instead of plotting individual lines, we plot a bar graph with slopes of best fit lines which can be interpreted as the rate of frost growth. The baseline case is plotted as a horizontal line and the remaining 8 cases are plotted as bar graphs. The numerical values of these rates are listed in \cref{tab:List of Simulations}. Effect of the four parameters on the rate of frost growth is as follows:
\begin{itemize}
	\item \textbf{Inlet velocity (UI)}: For velocities of [0.1, 0.3, 0.5] m/s, rates are [0.0629, 0.1043, 0.1294] g/min respectively. Higher velocity implies higher flow rate and thus, humid air brings more moisture content with it. Hence, frost is deposited faster at higher velocities. 
	\item \textbf{Inlet relative humidity (HI)}: For relative humidity of [40, 60, 80] \%, rates are [0.0322, 0.1043, 0.1648] g/min respectively. The amount of water vapor in air increases with relative humidity which causes higher frost deposition.
	\item \textbf{Inlet temperature (TI)}: For inlet temperatures of [5, 10, 15] $^\text{o}$C, rates are [0.0529, 0.1043, 0.1729] g/min respectively. Since we keep the relative humidity constant (60\%) for these 3 simulations, the water vapor in humid air content increases with inlet temperature and thus, we see higher rate of frost growth.
	\item \textbf{Surface temperature (TS)}: For surface temperatures of [-15, -10, -5] $^\text{o}$C, rates are [0.1312, 0.1043, 0.0628] g/min respectively. At lower surface temperature, air in the vicinity of the surface further cools down. Hence, more frost is deposited.
\end{itemize}
Contours of velocities, temperature, humidity and ice volume fraction are plotted for 9 cases in the following sections.
\subsection{Analysis of Baseline Case: B0} \label{Sec:Analysis of Baseline Case: B0}
\begin{figure}[H]
	\centering
	\begin{subfigure}[t]{0.49\textwidth}
		\includegraphics[width=\textwidth]{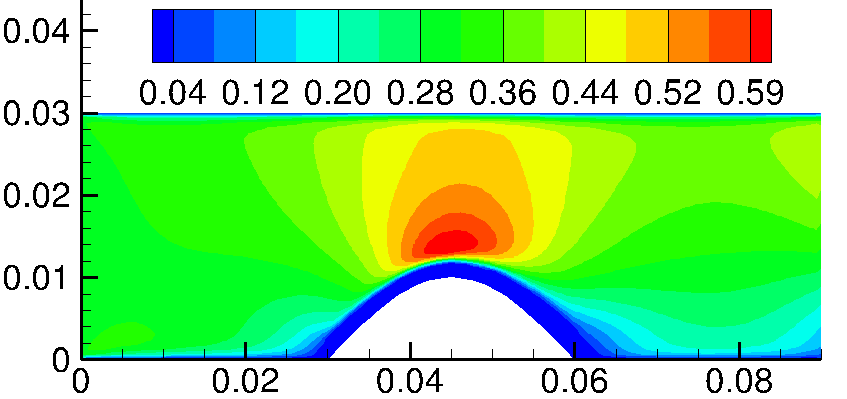}
		\caption{X Velocity (m/s)}
		\label{Fig:B0 Contour X-vel}
	\end{subfigure}
	\begin{subfigure}[t]{0.49\textwidth}
		\includegraphics[width=\textwidth]{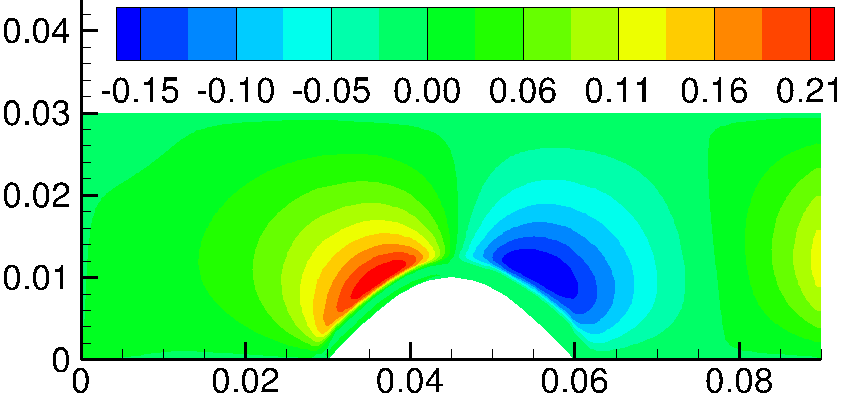}
		\caption{Y Velocity (m/s)}
		\label{Fig:B0 Contour Y-vel}
	\end{subfigure}
	\begin{subfigure}[t]{0.49\textwidth}
		\includegraphics[width=\textwidth]{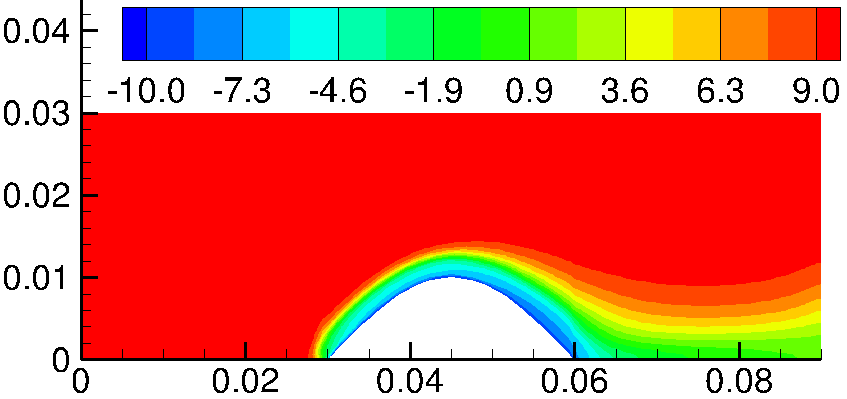}
		\caption{Temperature ($^\text{o}$C)}
		\label{Fig:B0 Contour T}
	\end{subfigure}
	\begin{subfigure}[t]{0.49\textwidth}
		\includegraphics[width=\textwidth]{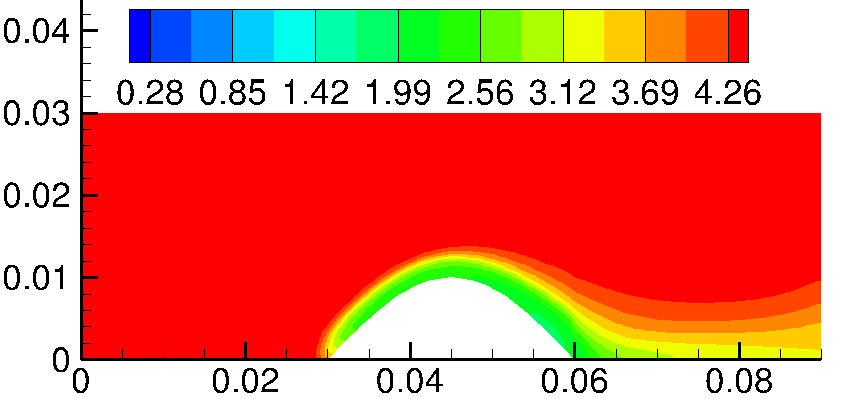}
		\caption{$Y$: Ratio of Water Vapor Mass to Humid Air Mass ($\text{g}_{\text{wv}}/\text{kg}_{\text{ha}}$)}
		\label{Fig:B0 Contour Y}
	\end{subfigure}
	\caption{Simulation B0: Contours After 2 Hours}
	\label{Fig:B0 Contour}
\end{figure}
\begin{figure}[H]
	\centering
	\begin{subfigure}[t]{0.49\textwidth}
		\includegraphics[width=\textwidth]{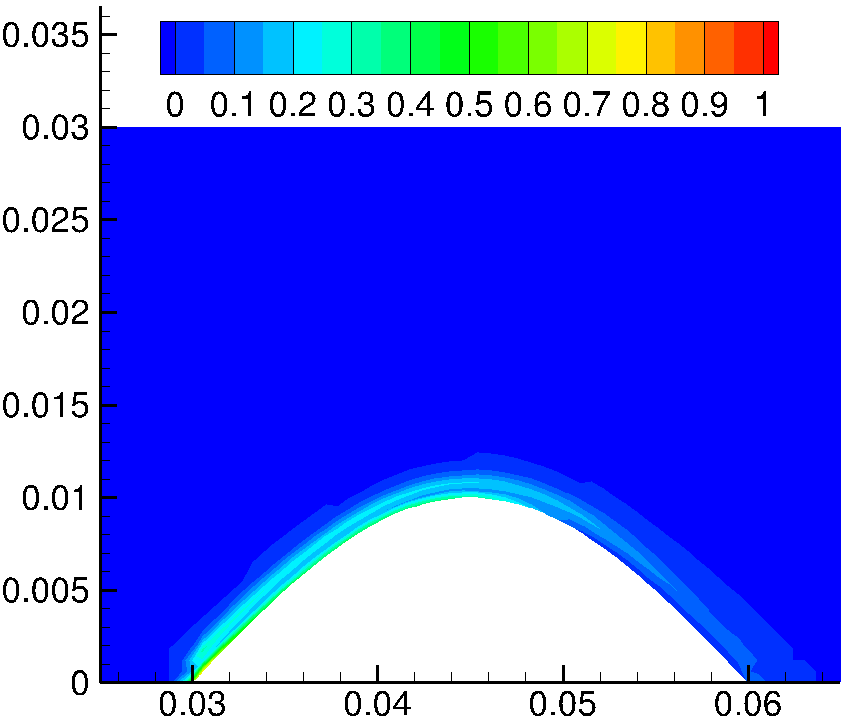}
		\caption{After 1 Hour}
		\label{Fig:B0 Contour alpha 1 hr}
	\end{subfigure}
	\begin{subfigure}[t]{0.49\textwidth}
		\includegraphics[width=\textwidth]{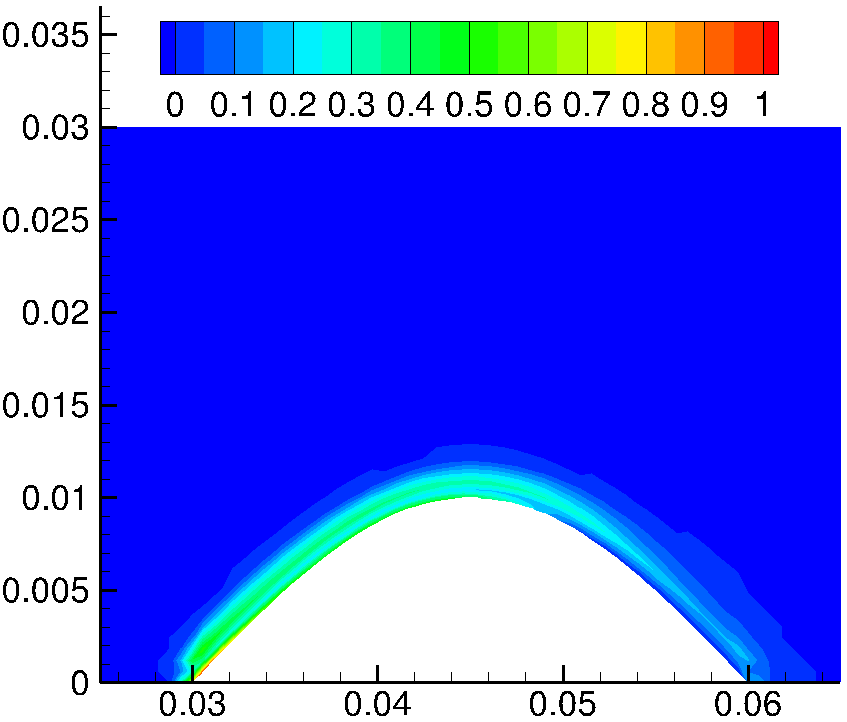}
		\caption{After 2 Hours}
		\label{Fig:B0 Contour alpha 2 hr}
	\end{subfigure}
	\caption{Simulation B0: Contours of Ice Volume Fraction ($\alpha_{ice}$)}
	\label{Fig:B0 Contour alpha}
\end{figure}

\Cref{Fig:B0 Contour} plots contours of velocities, temperature and humidity ratio for the baseline case after 2 hours of frost growth. Due to the sinusoidal surface, the flow turns in the upward direction. The flow also accelerates because of reduction in the cross--sectional area. Although the inlet velocity is 0.3 m/s, much higher velocities are observed in \cref{Fig:B0 Contour X-vel,Fig:B0 Contour Y-vel}. The sinusoidal surface is cooled down to $-10^\text{o}$C and thus, the humidity ratio reduces in it vicinity since the capacity of air to hold water vapor drops with temperature (\cref{Fig:B0 Contour T,Fig:B0 Contour Y}). A strong correlation between temperature and humidity ratio can be seen. All surfaces other than the sinusoidal surface are thermally insulated. Hence, the region upstream of the cold surface shows temperatures close to the inlet temperature. However, we see a much colder region downstream. This can be attributed to the convection term in the enthalpy \cref{Eq:enthalpy conservation}. \Cref{Fig:B0 Contour alpha} plots contours of ice volume fraction ($\alpha_{ice}$) after 1 and 2 hours. Near the surface, $\alpha_{ice}$ is close to unity since a layer of frost is deposited. In the region of humid air without ice, $\alpha_{ice}$ is zero. The rise in frost amount can be clearly seen from \cref{Fig:B0 Contour alpha 1 hr} to \cref{Fig:B0 Contour alpha 2 hr} as the frost thickness increases with time. We also observe that the upstream half of the sinusoidal surface has more frost deposited since humid air is incident on it first. By the time humid air reaches to the half region on the downstream side, the amount of water vapor it carries reduces and thus, lesser frost is deposited here. The frost region has higher porosity and thus, we see much lower velocities. This effect is achieved by the Darcy drag term in the momentum \cref{Eq:momentum x,Eq:momentum y}.

\subsection{Effect of Inlet Velocity: UI}

\begin{figure}[H]
	\centering
	\begin{subfigure}[t]{0.49\textwidth}
		\includegraphics[width=\textwidth]{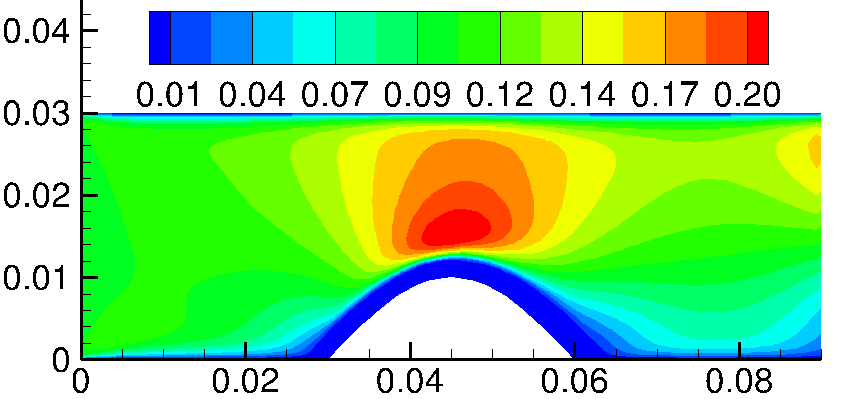}
		\caption{Simulation UI1: 0.1 m/s}
		\label{Fig:UI1 Contours X-vel}
	\end{subfigure}
	\begin{subfigure}[t]{0.49\textwidth}
		\includegraphics[width=\textwidth]{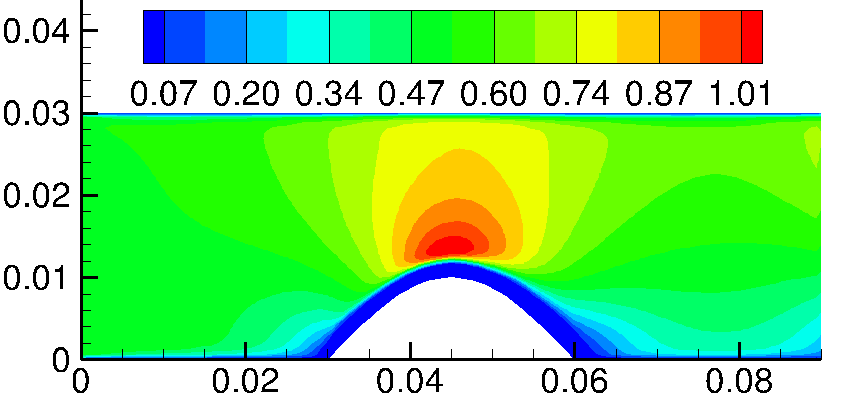}
		\caption{Simulation UI2: 0.5 m/s}
		\label{Fig:UI2 Contours X-vel}
	\end{subfigure}
	\caption{Simulations UI: Contours of X Velocity (m/s) After 2 Hours}
	\label{Fig:UI Contours X-vel}
\end{figure}
\begin{figure}[H]
	\centering
	\begin{subfigure}[t]{0.49\textwidth}
		\includegraphics[width=\textwidth]{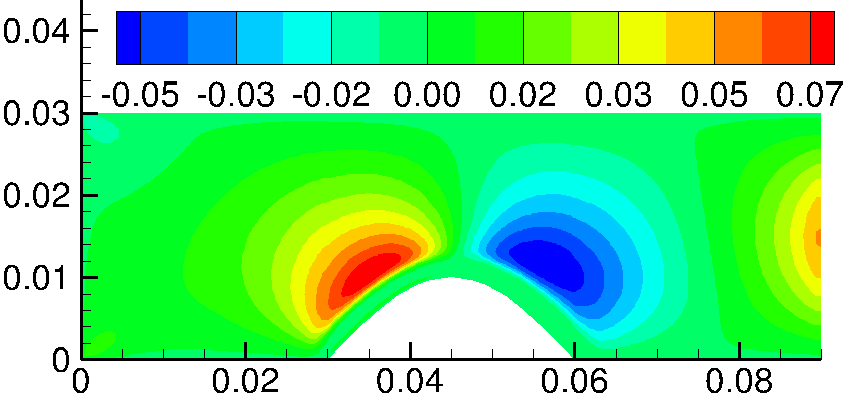}
		\caption{Simulation UI1: 0.1 m/s}
		\label{Fig:UI1 Contours Y-vel}
	\end{subfigure}
	\begin{subfigure}[t]{0.49\textwidth}
		\includegraphics[width=\textwidth]{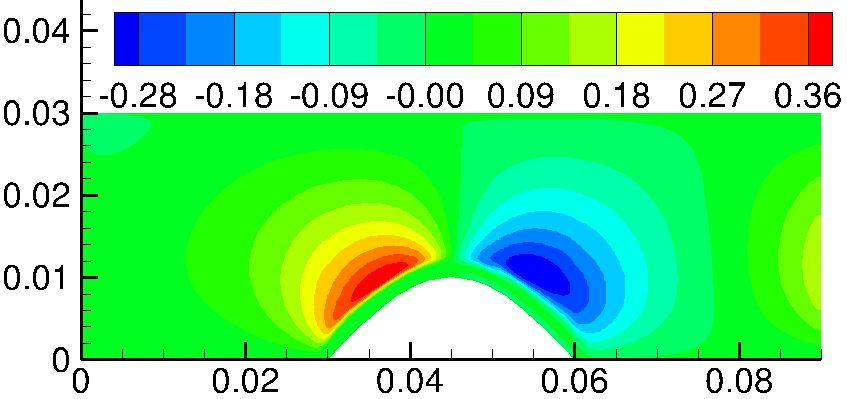}
		\caption{Simulation UI2: 0.5 m/s}
		\label{Fig:UI2 Contours Y-vel}
	\end{subfigure}
	\caption{Simulations UI: Contours of Y Velocity (m/s) After 2 Hours}
	\label{Fig:UI Contours Y-vel}
\end{figure}
\begin{figure}[H]
	\centering
	\begin{subfigure}[t]{0.49\textwidth}
		\includegraphics[width=\textwidth]{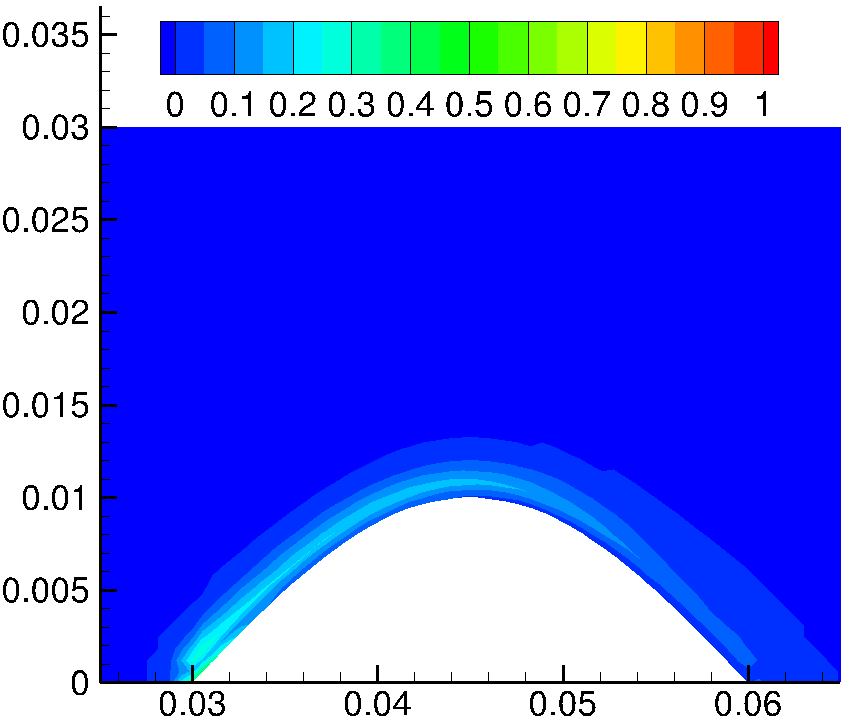}
		\caption{Simulation UI1: 0.1 m/s}
		\label{Fig:UI1 Contours alpha}
	\end{subfigure}
	\begin{subfigure}[t]{0.49\textwidth}
		\includegraphics[width=\textwidth]{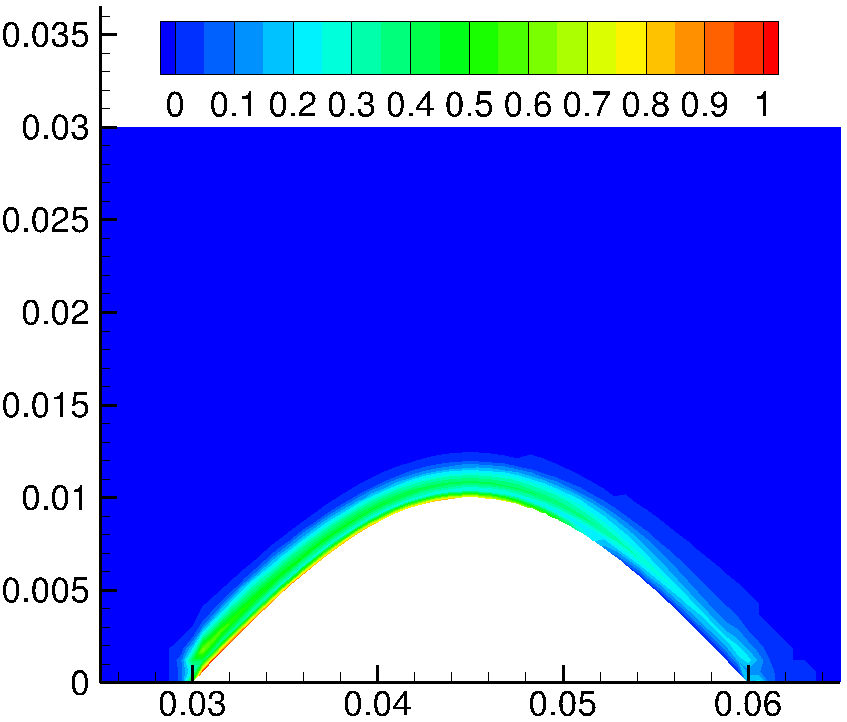}
		\caption{Simulation UI2: 0.5 m/s}
		\label{Fig:UI2 Contours alpha}
	\end{subfigure}
	\caption{Simulations UI: Contours of Ice Volume Fraction ($\alpha_{ice}$) After 2 Hours}
	\label{Fig:UI Contours alpha}
\end{figure}

In the following sections we plot  contours selectively in the interest of brevity. They can be compared with plots in \cref{Sec:Analysis of Baseline Case: B0} for better understanding. \Cref{Fig:UI Contours X-vel,Fig:UI Contours Y-vel} plot contours of velocity components for inlet velocity lower and higher than the baseline case. We see similar trends for both cases but with higher values. The trends in the volume fraction of ice are significantly different depending on the inlet velocity (\cref{Fig:UI Contours alpha}). As mentioned before, at higher inlet velocity, the water vapor content in the air increases, thus causing higher frost deposition.

\subsection{Effect of Inlet Humidity: HI}
\begin{figure}[H]
	\centering
	\begin{subfigure}[t]{0.49\textwidth}
		\includegraphics[width=\textwidth]{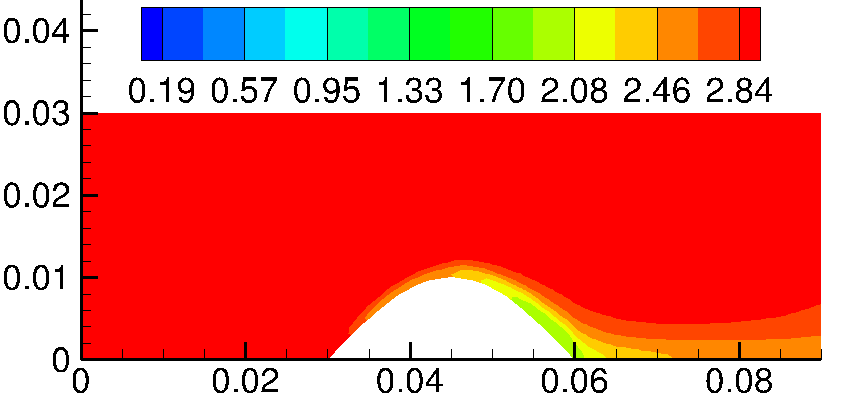}
		\caption{Simulation HI1: 40\%}
		\label{Fig:HI1 Contours Y}
	\end{subfigure}
	\begin{subfigure}[t]{0.49\textwidth}
		\includegraphics[width=\textwidth]{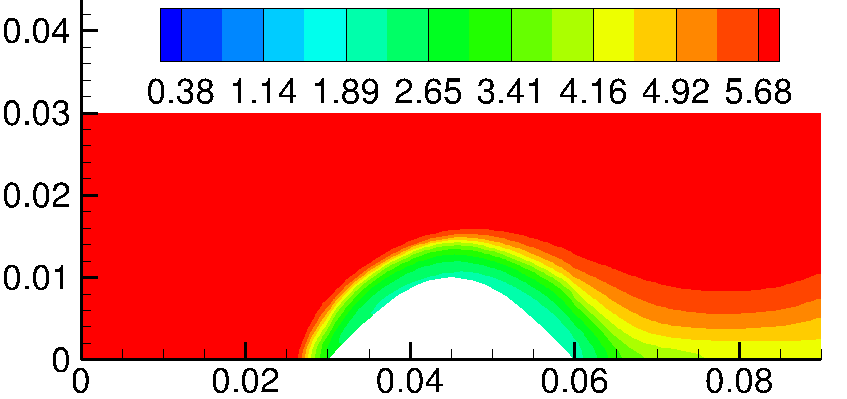}
		\caption{Simulation HI2: 80\%}
		\label{Fig:HI2 Contours Y}
	\end{subfigure}
	\caption{Simulations HI: Contours of $Y$: Ratio of Water Vapor Mass to Humid Air Mass ($\text{g}_{\text{wv}}/\text{kg}_{\text{ha}}$) After 2 Hours}
	\label{Fig:HI Contours Y}
\end{figure}

As expected, for the case with 80\% humidity (\cref{Fig:HI2 Contours Y}), we see the humidity ratio is almost twice compared to the case with 40\% inlet humidity (\cref{Fig:HI1 Contours Y}). The amount of frost deposited also increases significantly and a thick layer of low humidity air is observed in the vicinity of the cold surface for the case HI2.

\begin{figure}[H]
	\centering
	\begin{subfigure}[t]{0.49\textwidth}
		\includegraphics[width=\textwidth]{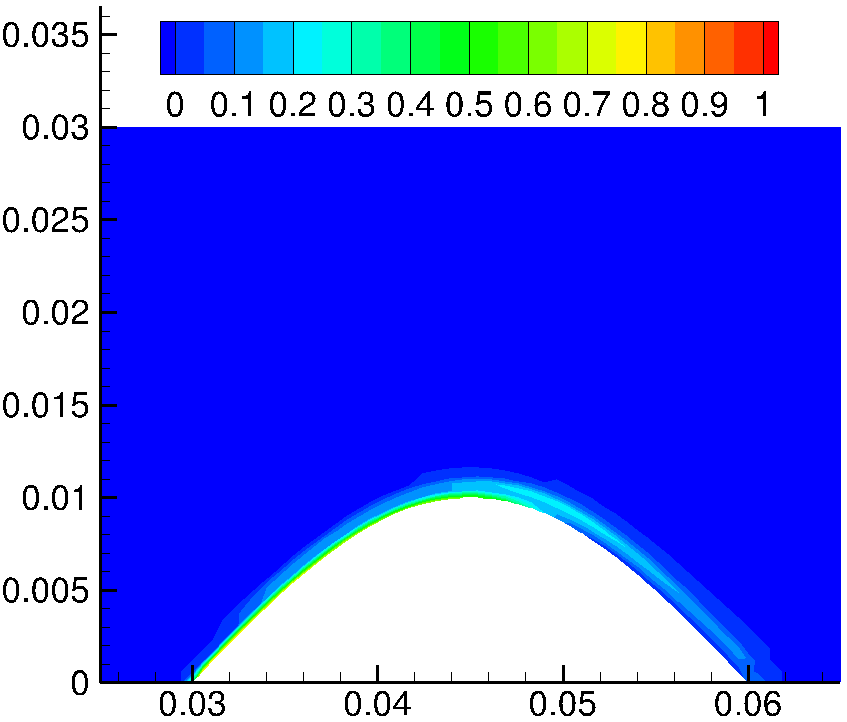}
		\caption{Simulation HI1: 40\%}
		\label{Fig:HI1 Contours alpha}
	\end{subfigure}
	\begin{subfigure}[t]{0.49\textwidth}
		\includegraphics[width=\textwidth]{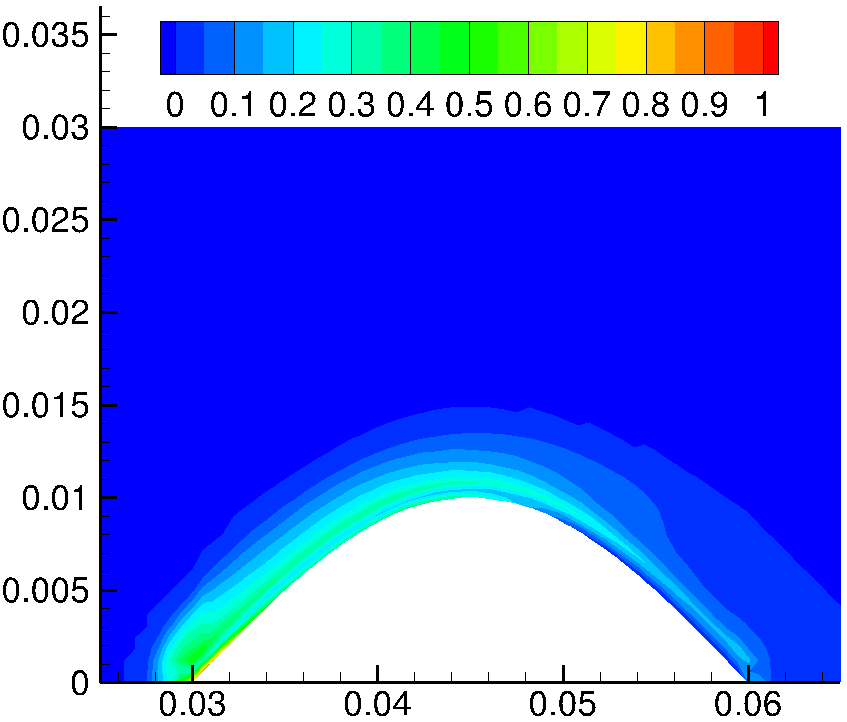}
		\caption{Simulation HI2: 80\%}
		\label{Fig:HI2 Contours alpha}
	\end{subfigure}
	\caption{Simulations HI: Contours of Ice Volume Fraction ($\alpha_{ice}$) After 2 Hours}
	\label{Fig:HI Contours alpha}
\end{figure}

\subsection{Effect of Inlet Temperature: TI}

\begin{figure}[H]
	\centering
	\begin{subfigure}[t]{0.49\textwidth}
		\includegraphics[width=\textwidth]{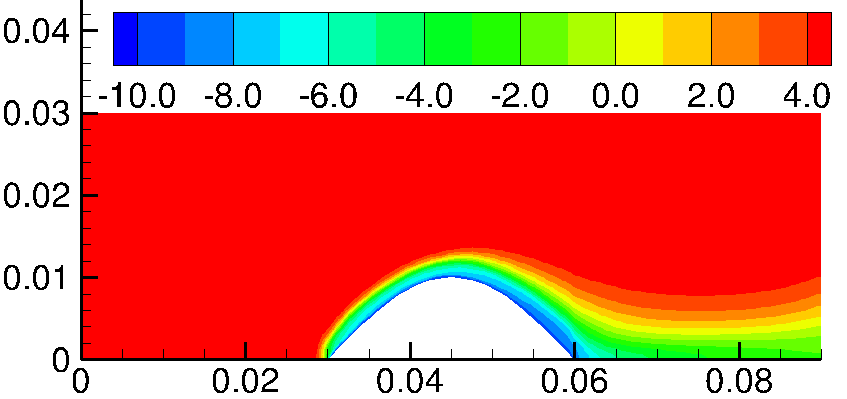}
		\caption{Simulation TI1: 5$^\text{o}$C}
		\label{Fig:TI1 Contours T}
	\end{subfigure}
	\begin{subfigure}[t]{0.49\textwidth}
		\includegraphics[width=\textwidth]{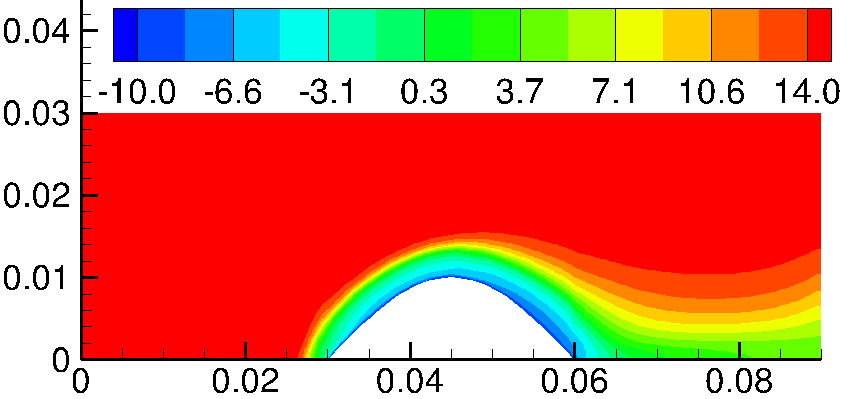}
		\caption{Simulation TI2: 15$^\text{o}$C}
		\label{Fig:TI2 Contours T}
	\end{subfigure}
	\caption{Simulations TI: Contours of Temperature ($^\text{o}$C) After 2 Hours}
	\label{Fig:TI Contours T}
\end{figure}
The patterns of temperature are similar for both the cases except the value of high temperature of 15$^\text{o}$C for the case TI2. However, we see much higher frost growth for TI2. This is because the humid air with higher inlet temperature at the same relative humidity of 60\% carries more water vapor.
\begin{figure}[H]
	\centering
	\begin{subfigure}[t]{0.49\textwidth}
		\includegraphics[width=\textwidth]{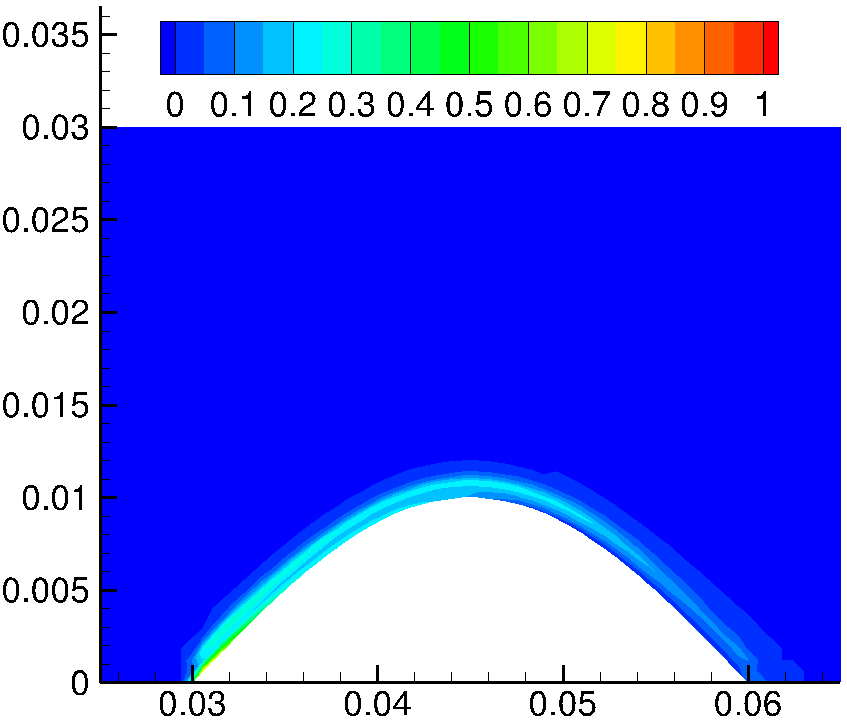}
		\caption{Simulation TI1: 5$^\text{o}$C}
		\label{Fig:TI1 Contours alpha}
	\end{subfigure}
	\begin{subfigure}[t]{0.49\textwidth}
		\includegraphics[width=\textwidth]{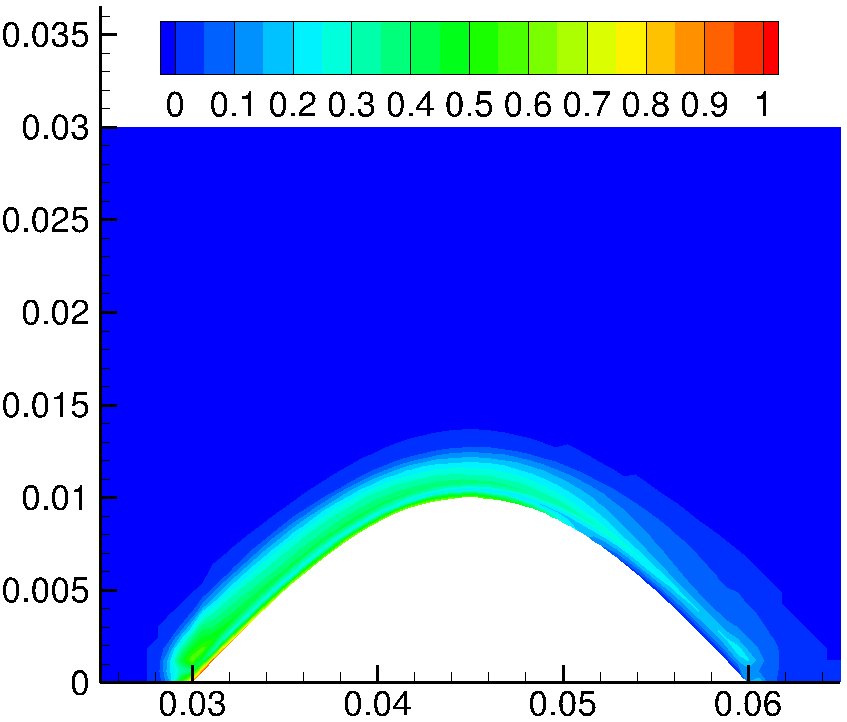}
		\caption{Simulation TI2: 15$^\text{o}$C}
		\label{Fig:TI2 Contours alpha}
	\end{subfigure}
	\caption{Simulations TI: Contours of Ice Volume Fraction ($\alpha_{ice}$) After 2 Hours}
	\label{Fig:TI Contours alpha}
\end{figure}

\subsection{Effect of Surface Temperature: TS}

\begin{figure}[H]
	\centering
	\begin{subfigure}[t]{0.49\textwidth}
		\includegraphics[width=\textwidth]{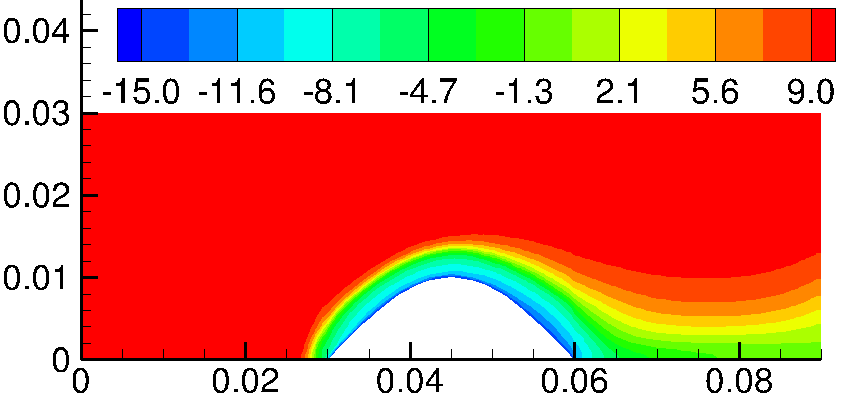}
		\caption{Simulation TS1: --15$^\text{o}$C}
		\label{Fig:TS1 Contours T}
	\end{subfigure}
	\begin{subfigure}[t]{0.49\textwidth}
		\includegraphics[width=\textwidth]{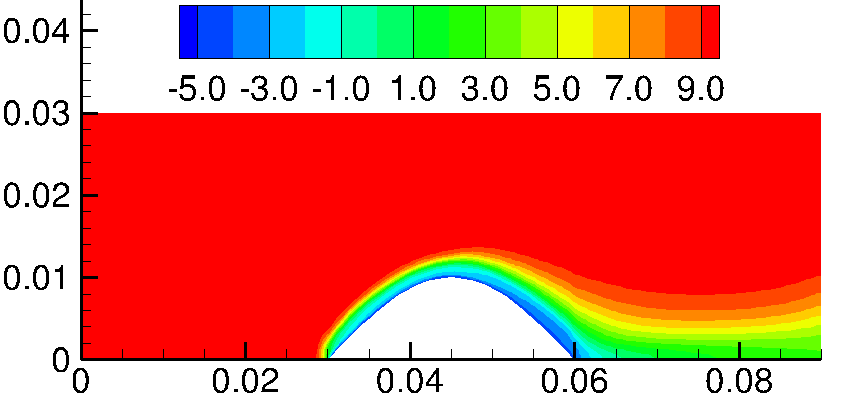}
		\caption{Simulation TS2: --5$^\text{o}$C}
		\label{Fig:TS2 Contours T}
	\end{subfigure}
	\caption{Simulations TS: Contours of Temperature ($^\text{o}$C) After 2 Hours}
	\label{Fig:TS Contours T}
\end{figure}
\begin{figure}[H]
	\centering
	\begin{subfigure}[t]{0.49\textwidth}
		\includegraphics[width=\textwidth]{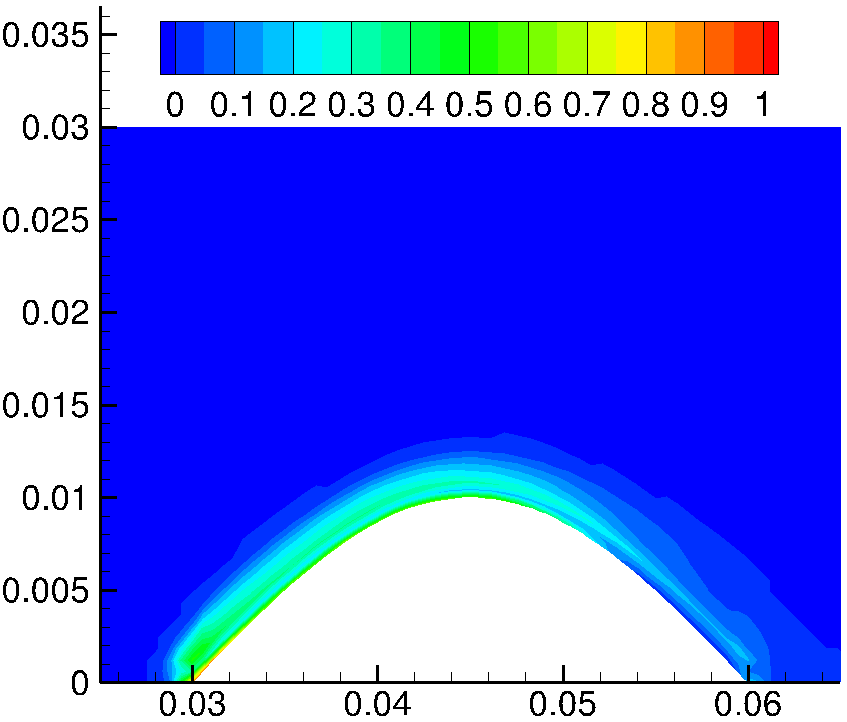}
		\caption{Simulation TS1: --15$^\text{o}$C}
		\label{Fig:TS1 Contours alpha}
	\end{subfigure}
	\begin{subfigure}[t]{0.49\textwidth}
		\includegraphics[width=\textwidth]{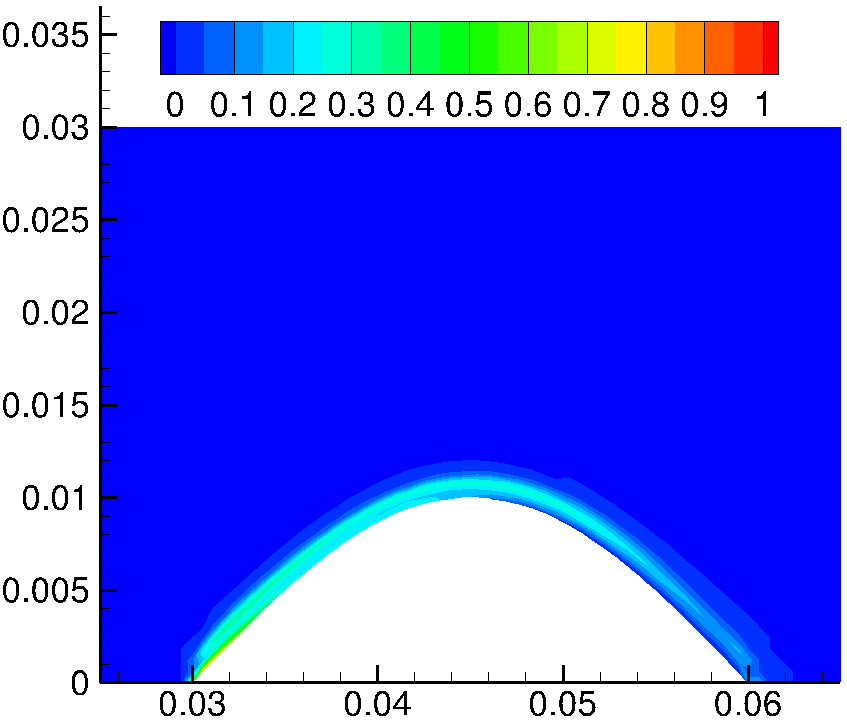}
		\caption{Simulation TS2: --5$^\text{o}$C}
		\label{Fig:TS2 Contours alpha}
	\end{subfigure}
	\caption{Simulations TS: Contours of Ice Volume Fraction ($\alpha_{ice}$) After 2 Hours}
	\label{Fig:TS Contours alpha}
\end{figure}

Case TS1 has a lower surface temperature than TS2. A thicker region of cold air is thus formed around the sinusoidal surface (\cref{Fig:TS Contours T}). As air cools down, its capacity to hold water vapor decreases. The extra water vapor desublimates to frost. Thus, TS1 has a significantly thicker frost layer than TS2 (\cref{Fig:TS Contours alpha}).

\subsection{Effect of Surface Wettability (CA)}
\begin{figure}[H]
	\centering
	\begin{subfigure}[t]{0.32\textwidth}
		\includegraphics[width=\textwidth]{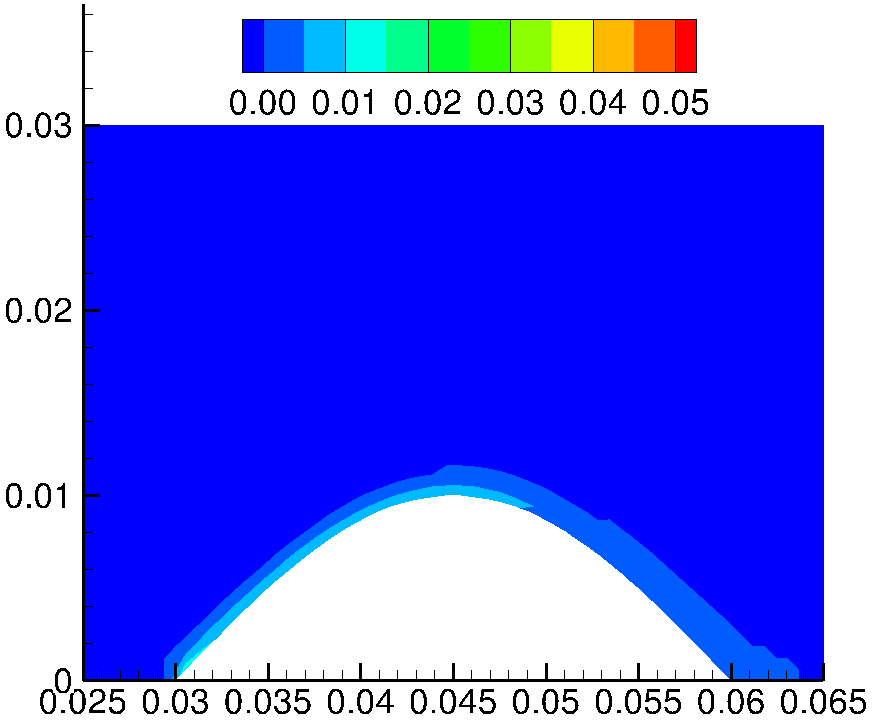}
		\caption{Simulation CA1: 30$^\text{o}$ Contact Angle}
		\label{Fig:CA1 Contours mass ice 5 minutes}
	\end{subfigure}
	\begin{subfigure}[t]{0.32\textwidth}
		\includegraphics[width=\textwidth]{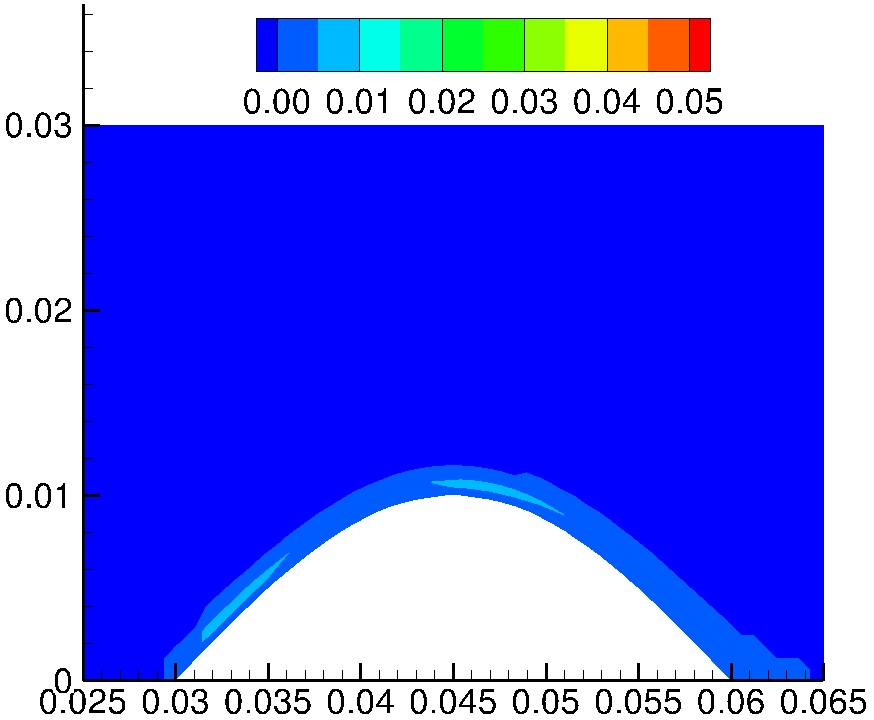}
		\caption{Simulation B0: 90$^\text{o}$ Contact Angle}
		\label{Fig:B0 Contours mass ice 5 minutes}
	\end{subfigure}
	\begin{subfigure}[t]{0.32\textwidth}
		\includegraphics[width=\textwidth]{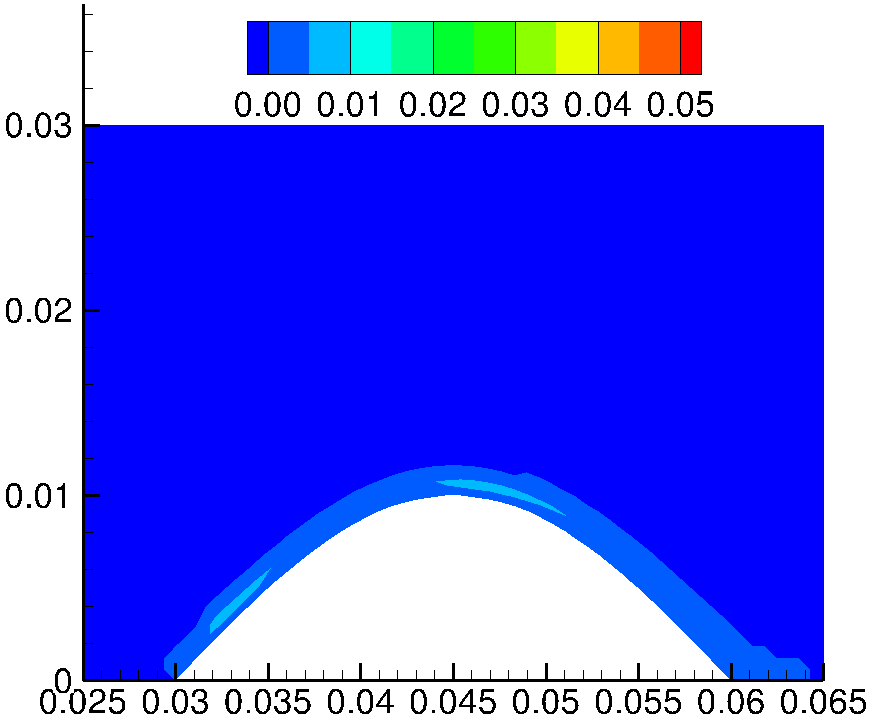}
		\caption{Simulation CA2: 150$^\text{o}$ Contact Angle}
		\label{Fig:CA2 Contours mass ice 5 minutes}
	\end{subfigure}
	\caption{Simulation CA: Contours of Frost Mass After 5 Minutes}
	\label{Fig:CA Contours mass ice 5 minutes}
\end{figure}
\begin{figure}[H]
	\centering
	\begin{subfigure}[t]{0.32\textwidth}
		\includegraphics[width=\textwidth]{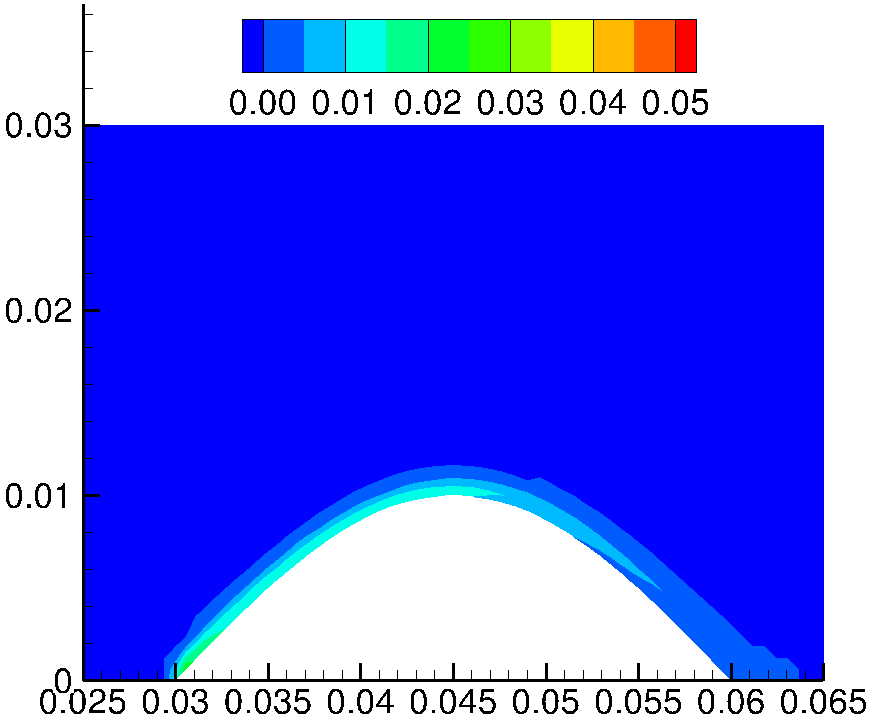}
		\caption{Simulation CA1: 30v Contact Angle}
		\label{Fig:CA1 Contours mass ice 10 minutes}
	\end{subfigure}
	\begin{subfigure}[t]{0.32\textwidth}
		\includegraphics[width=\textwidth]{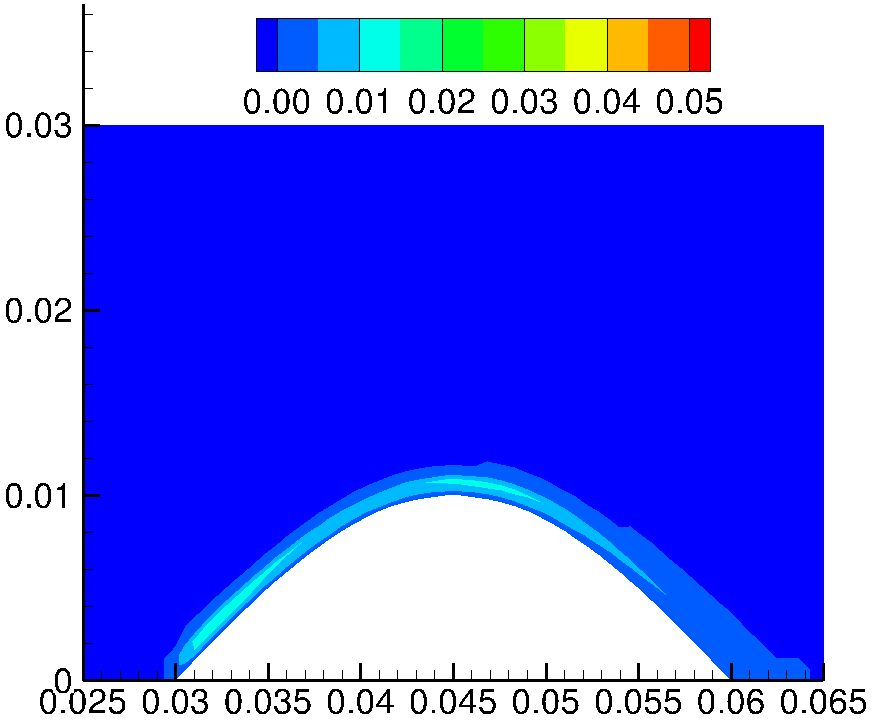}
		\caption{Simulation B0: 90$^\text{o}$ Contact Angle}
		\label{Fig:B0 Contours mass ice 10 minutes}
	\end{subfigure}
	\begin{subfigure}[t]{0.32\textwidth}
		\includegraphics[width=\textwidth]{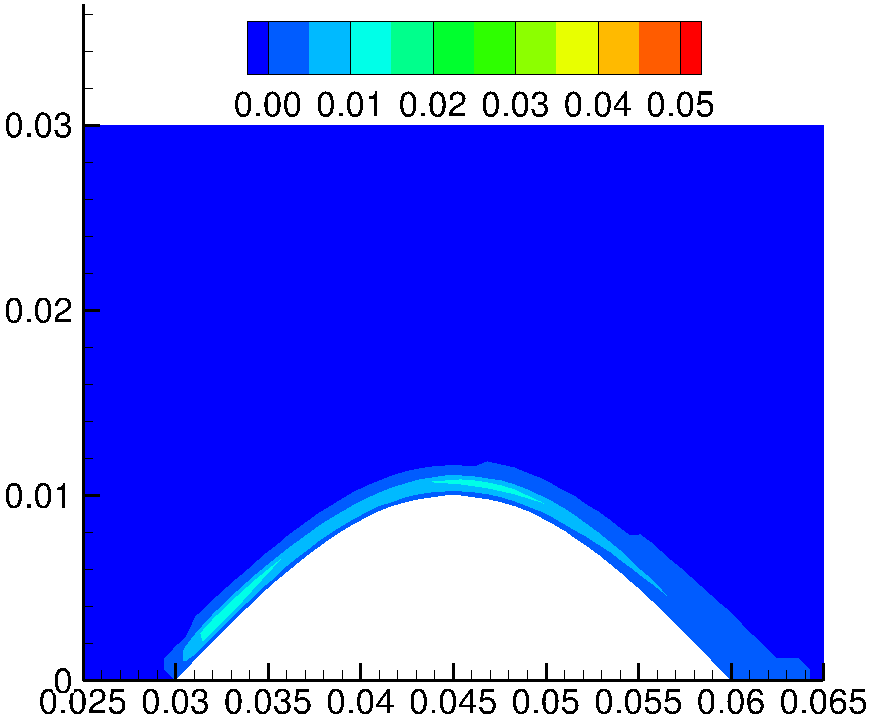}
		\caption{Simulation CA2: 150$^\text{o}$ Contact Angle}
		\label{Fig:CA2 Contours mass ice 10 minutes}
	\end{subfigure}
	\caption{Simulation CA: Contours of Frost Mass After 10 Minutes}
	\label{Fig:CA Contours mass ice 10 minutes}
\end{figure}
The surface wettability effect on frost growth is compared under surface contact angle of 30$^\text{o}$ (hydrophilic), 90$^\text{o}$ and 150$^\text{o}$ (superhydrophobic) shown in the frost mass contour in \cref{Fig:CA Contours mass ice 10 minutes,Fig:CA Contours mass ice 5 minutes}. It can be noted that frost mass is higher and more uniformly distributed on surface with lower contact angle, especially at the first 5 minutes, frost is scattered on the leading edge and the top of the sine wave of the hydrophobic surface. It can be explained by the supersaturation degree for different surfaces. The supersaturation required for nucleation on surface with low contact angle is lower than that with high contact angle, which allows the nucleation occurs earlier and results in more deposition of the frost during the earlier frosting process. It seems the surface wettability mainly contributes to the initial frost growth and the effect is gradually diminished with the following buildup of the frost layer as shown after 10 minutes in \cref{Fig:CA Contours mass ice 10 minutes}.
\section{Conclusions}
In this work, a numerical model has been developed to predict frost growth under varying surface temperature, wettability, air temperature, humidity and flow velocity. The mixture model has been adapted to implicitly track the frost--air interface. This approach is beneficial since the exact interface conditions may not be available for frost growth on complex geometries. Empirical relations are used to estimate spatial variations of viscosity, thermal conductivity and diffusion resistance factor. Local variations in other properties such as density and specific heat capacity is also considered. The semi--implicit pressure splitting method has been used to solve the momentum equations coupled with mass conservation and energy equations. This approach allows a convection Courant number over unity thus, improving computational efficiency by reducing the total number of timesteps. The governing equations are discretized on unstructured grids to model complex geometries. The model is first validated using experimental data in the literature. A good agreement of frost thickness with the experimental measurement shows that the model is accurate.
\par The frost distribution has been demonstrated on a two--dimensional sinusoidal surface. A detailed analysis of temperature, velocity and water vapor content in the air and frost regions is presented. Higher air velocity, temperature or humidity gets more water vapor, and thus, the rate of frost growth increases. Similarly, reduction in surface temperature causes a higher frost deposition due to enhanced cooling. Higher surface wettability generates more frost during the early stage of frosting. To the best of our knowledge, a mixture model with semi--implicit pressure splitting approach implemented on an unstructured grid has not been reported in the literature for modeling of frost growth. The accuracy and computational efficiency are the salient features of this approach. In future, we plan to apply the method to estimate frost growth on various heat exchanger surfaces with complex geometries.

\section*{Acknowledgments}
The authors are grateful for the financial support from the Air Conditioning and Refrigeration Center (ACRC) at the University of Illinois at Urbana--Champaign.

\bibliography{References}

\begin{thebibliography}{69}
\expandafter\ifx\csname natexlab\endcsname\relax\def\natexlab#1{#1}\fi
\providecommand{\url}[1]{\texttt{#1}}
\providecommand{\href}[2]{#2}
\providecommand{\path}[1]{#1}
\providecommand{\DOIprefix}{doi:}
\providecommand{\ArXivprefix}{arXiv:}
\providecommand{\URLprefix}{URL: }
\providecommand{\Pubmedprefix}{pmid:}
\providecommand{\doi}[1]{\href{http://dx.doi.org/#1}{\path{#1}}}
\providecommand{\Pubmed}[1]{\href{pmid:#1}{\path{#1}}}
\providecommand{\bibinfo}[2]{#2}
\ifx\xfnm\relax \def\xfnm[#1]{\unskip,\space#1}\fi
\bibitem[{Shen and Wang(2019)}]{shen2019real}
\bibinfo{author}{Y.~Shen}, \bibinfo{author}{X.~Wang},
\newblock \bibinfo{title}{Real-time frost porosity detection using capacitance
  sensing approach},
\newblock \bibinfo{journal}{International Journal of Heat and Mass Transfer}
  \bibinfo{volume}{134} (\bibinfo{year}{2019}) \bibinfo{pages}{1171--1179}.
\bibitem[{Shen and Wang(2020)}]{shen2020condensation1}
\bibinfo{author}{Y.~Shen}, \bibinfo{author}{S.~Wang},
\newblock \bibinfo{title}{Condensation frosting detection and characterization
  using a capacitance sensing approach},
\newblock \bibinfo{journal}{International Journal of Heat and Mass Transfer}
  \bibinfo{volume}{147} (\bibinfo{year}{2020}) \bibinfo{pages}{118968}.
\bibitem[{Shen et~al.(2020)Shen, Zou, and Wang}]{shen2020condensation2}
\bibinfo{author}{Y.~Shen}, \bibinfo{author}{H.~Zou}, \bibinfo{author}{S.~Wang},
\newblock \bibinfo{title}{Condensation frosting on micropillar surfaces--effect
  of microscale roughness on ice propagation},
\newblock \bibinfo{journal}{Langmuir} \bibinfo{volume}{36}
  (\bibinfo{year}{2020}) \bibinfo{pages}{13563--13574}.
\bibitem[{Kandula(2011)}]{kandula2011frost}
\bibinfo{author}{M.~Kandula},
\newblock \bibinfo{title}{Frost growth and densification in laminar flow over
  flat surfaces},
\newblock \bibinfo{journal}{International Journal of Heat and Mass Transfer}
  \bibinfo{volume}{54} (\bibinfo{year}{2011}) \bibinfo{pages}{3719--3731}.
\bibitem[{Hermes et~al.(2009)Hermes, Piucco, Barbosa~Jr, and
  Melo}]{hermes2009study}
\bibinfo{author}{C.~J. Hermes}, \bibinfo{author}{R.~O. Piucco},
  \bibinfo{author}{J.~R. Barbosa~Jr}, \bibinfo{author}{C.~Melo},
\newblock \bibinfo{title}{A study of frost growth and densification on flat
  surfaces},
\newblock \bibinfo{journal}{Experimental Thermal and Fluid Science}
  \bibinfo{volume}{33} (\bibinfo{year}{2009}) \bibinfo{pages}{371--379}.
\bibitem[{Yun et~al.(2002)Yun, Kim, and Min}]{yun2002modeling}
\bibinfo{author}{R.~Yun}, \bibinfo{author}{Y.~Kim}, \bibinfo{author}{M.-k.
  Min},
\newblock \bibinfo{title}{Modeling of frost growth and frost properties with
  airflow over a flat plate},
\newblock \bibinfo{journal}{International Journal of Refrigeration}
  \bibinfo{volume}{25} (\bibinfo{year}{2002}) \bibinfo{pages}{362--371}.
\bibitem[{Rabbi et~al.(2021)Rabbi, Boyina, Su, Sett, Thamban, Shahane, Wang,
  and Miljkovic}]{rabbi2021wettability}
\bibinfo{author}{K.~F. Rabbi}, \bibinfo{author}{K.~S. Boyina},
  \bibinfo{author}{W.~Su}, \bibinfo{author}{S.~Sett},
  \bibinfo{author}{A.~Thamban}, \bibinfo{author}{S.~Shahane},
  \bibinfo{author}{S.~Wang}, \bibinfo{author}{N.~Miljkovic},
\newblock \bibinfo{title}{Wettability-defined frosting dynamics between plane
  fins in quiescent air},
\newblock \bibinfo{journal}{International Journal of Heat and Mass Transfer}
  \bibinfo{volume}{164} (\bibinfo{year}{2021}) \bibinfo{pages}{120563}.
\bibitem[{Wang et~al.(2016)Wang, Xiong, Lu, Pan, Wang, Deng, and
  Shi}]{wang2016design}
\bibinfo{author}{N.~Wang}, \bibinfo{author}{D.~Xiong}, \bibinfo{author}{Y.~Lu},
  \bibinfo{author}{S.~Pan}, \bibinfo{author}{K.~Wang},
  \bibinfo{author}{Y.~Deng}, \bibinfo{author}{Y.~Shi},
\newblock \bibinfo{title}{Design and fabrication of the lyophobic slippery
  surface and its application in anti-icing},
\newblock \bibinfo{journal}{The Journal of Physical Chemistry C}
  \bibinfo{volume}{120} (\bibinfo{year}{2016}) \bibinfo{pages}{11054--11059}.
\bibitem[{Liu et~al.(2016)Liu, Zhang, Tao, Zhao, Li, Zhu, and
  Yuan}]{liu2016strategies}
\bibinfo{author}{B.~Liu}, \bibinfo{author}{K.~Zhang}, \bibinfo{author}{C.~Tao},
  \bibinfo{author}{Y.~Zhao}, \bibinfo{author}{X.~Li}, \bibinfo{author}{K.~Zhu},
  \bibinfo{author}{X.~Yuan},
\newblock \bibinfo{title}{Strategies for anti-icing: low surface energy or
  liquid-infused?},
\newblock \bibinfo{journal}{RSC advances} \bibinfo{volume}{6}
  (\bibinfo{year}{2016}) \bibinfo{pages}{70251--70260}.
\bibitem[{Chu et~al.(2018)Chu, Wen, and Wu}]{chu2018frost}
\bibinfo{author}{F.~Chu}, \bibinfo{author}{D.~Wen}, \bibinfo{author}{X.~Wu},
\newblock \bibinfo{title}{Frost self-removal mechanism during defrosting on
  vertical superhydrophobic surfaces: Peeling off or jumping off},
\newblock \bibinfo{journal}{Langmuir} \bibinfo{volume}{34}
  (\bibinfo{year}{2018}) \bibinfo{pages}{14562--14569}.
\bibitem[{Cheng and Wu(2003)}]{cheng2003observations}
\bibinfo{author}{C.-H. Cheng}, \bibinfo{author}{K.-H. Wu},
\newblock \bibinfo{title}{Observations of early-stage frost formation on a cold
  plate in atmospheric air flow},
\newblock \bibinfo{journal}{J. Heat Transfer} \bibinfo{volume}{125}
  (\bibinfo{year}{2003}) \bibinfo{pages}{95--102}.
\bibitem[{Lee et~al.(2003)Lee, Jhee, and Yang}]{lee2003prediction}
\bibinfo{author}{K.~Lee}, \bibinfo{author}{S.~Jhee}, \bibinfo{author}{D.~Yang},
\newblock \bibinfo{title}{Prediction of the frost formation on a cold flat
  surface},
\newblock \bibinfo{journal}{International Journal of Heat and Mass Transfer}
  \bibinfo{volume}{46} (\bibinfo{year}{2003}) \bibinfo{pages}{3789--3796}.
\bibitem[{Niroomand et~al.(2019)Niroomand, Fauchoux, and
  Simonson}]{niroomand2019experimental}
\bibinfo{author}{S.~Niroomand}, \bibinfo{author}{M.~Fauchoux},
  \bibinfo{author}{C.~Simonson},
\newblock \bibinfo{title}{Experimental characterization of frost growth on a
  horizontal plate under natural convection},
\newblock \bibinfo{journal}{Journal of Thermal Science and Engineering
  Applications} \bibinfo{volume}{11} (\bibinfo{year}{2019}).
\bibitem[{Chen et~al.(2019)Chen, Deng, Zhang, and Yan}]{chen2019simulation}
\bibinfo{author}{G.~Chen}, \bibinfo{author}{X.~Deng},
  \bibinfo{author}{G.~Zhang}, \bibinfo{author}{X.~Yan},
\newblock \bibinfo{title}{Simulation of frost growth and densification on
  horizontal plates with supersaturated interface condition},
\newblock \bibinfo{journal}{International Journal of Heat and Mass Transfer}
  \bibinfo{volume}{133} (\bibinfo{year}{2019}) \bibinfo{pages}{426--434}.
\bibitem[{Yang and Lee(2005)}]{yang2005modeling}
\bibinfo{author}{D.~Yang}, \bibinfo{author}{K.~Lee},
\newblock \bibinfo{title}{Modeling of frosting behavior on a cold plate},
\newblock \bibinfo{journal}{International Journal of Refrigeration}
  \bibinfo{volume}{28} (\bibinfo{year}{2005}) \bibinfo{pages}{396--402}.
\bibitem[{Cui et~al.(2011)Cui, Li, Liu, and Jiang}]{cui2011new_ate}
\bibinfo{author}{J.~Cui}, \bibinfo{author}{W.~Li}, \bibinfo{author}{Y.~Liu},
  \bibinfo{author}{Z.~Jiang},
\newblock \bibinfo{title}{A new time-and space-dependent model for predicting
  frost formation},
\newblock \bibinfo{journal}{Applied Thermal Engineering} \bibinfo{volume}{31}
  (\bibinfo{year}{2011}) \bibinfo{pages}{447--457}.
\bibitem[{Wu et~al.(2016)Wu, Ma, Chu, and Hu}]{wu2016phase}
\bibinfo{author}{X.~Wu}, \bibinfo{author}{Q.~Ma}, \bibinfo{author}{F.~Chu},
  \bibinfo{author}{S.~Hu},
\newblock \bibinfo{title}{Phase change mass transfer model for frost growth and
  densification},
\newblock \bibinfo{journal}{International Journal of Heat and Mass Transfer}
  \bibinfo{volume}{96} (\bibinfo{year}{2016}) \bibinfo{pages}{11--19}.
\bibitem[{Wu et~al.(2017)Wu, Chu, and Ma}]{wu2017frosting}
\bibinfo{author}{X.~Wu}, \bibinfo{author}{F.~Chu}, \bibinfo{author}{Q.~Ma},
\newblock \bibinfo{title}{Frosting model based on phase change driving force},
\newblock \bibinfo{journal}{International Journal of Heat and Mass Transfer}
  \bibinfo{volume}{110} (\bibinfo{year}{2017}) \bibinfo{pages}{760--767}.
\bibitem[{Cui et~al.(2011)Cui, Li, Liu, and Zhao}]{cui2011new_ijhff}
\bibinfo{author}{J.~Cui}, \bibinfo{author}{W.~Li}, \bibinfo{author}{Y.~Liu},
  \bibinfo{author}{Y.~Zhao},
\newblock \bibinfo{title}{A new model for predicting performance of
  fin-and-tube heat exchanger under frost condition},
\newblock \bibinfo{journal}{International Journal of Heat and Fluid Flow}
  \bibinfo{volume}{32} (\bibinfo{year}{2011}) \bibinfo{pages}{249--260}.
\bibitem[{Loyola et~al.(2014)Loyola, Nascimento~Jr, and
  Hermes}]{loyola2014modeling}
\bibinfo{author}{F.~Loyola}, \bibinfo{author}{V.~Nascimento~Jr},
  \bibinfo{author}{C.~Hermes},
\newblock \bibinfo{title}{Modeling of frost build-up on parallel-plate channels
  under supersaturated air-frost interface conditions},
\newblock \bibinfo{journal}{International Journal of Heat and Mass Transfer}
  \bibinfo{volume}{79} (\bibinfo{year}{2014}) \bibinfo{pages}{790--795}.
\bibitem[{El~Cheikh and Jacobi(2014)}]{el2014mathematical}
\bibinfo{author}{A.~El~Cheikh}, \bibinfo{author}{A.~Jacobi},
\newblock \bibinfo{title}{A mathematical model for frost growth and
  densification on flat surfaces},
\newblock \bibinfo{journal}{International Journal of Heat and Mass Transfer}
  \bibinfo{volume}{77} (\bibinfo{year}{2014}) \bibinfo{pages}{604--611}.
\bibitem[{Kim et~al.(2015)Kim, Kim, and Lee}]{kim2015frosting}
\bibinfo{author}{D.~Kim}, \bibinfo{author}{C.~Kim}, \bibinfo{author}{K.~Lee},
\newblock \bibinfo{title}{Frosting model for predicting macroscopic and local
  frost behaviors on a cold plate},
\newblock \bibinfo{journal}{International Journal of Heat and Mass Transfer}
  \bibinfo{volume}{82} (\bibinfo{year}{2015}) \bibinfo{pages}{135--142}.
\bibitem[{Armengol et~al.(2016)Armengol, Salinas, Xaman, and
  Ismail}]{armengol2016modeling}
\bibinfo{author}{J.~Armengol}, \bibinfo{author}{C.~Salinas},
  \bibinfo{author}{J.~Xaman}, \bibinfo{author}{K.~Ismail},
\newblock \bibinfo{title}{Modeling of frost formation over parallel cold plates
  considering a two-dimensional growth rate},
\newblock \bibinfo{journal}{International Journal of Thermal Sciences}
  \bibinfo{volume}{104} (\bibinfo{year}{2016}) \bibinfo{pages}{245--256}.
\bibitem[{Shahane et~al.(2019)Shahane, Aluru, Ferreira, Kapoor, and
  Vanka}]{shahane2019finite}
\bibinfo{author}{S.~Shahane}, \bibinfo{author}{N.~Aluru},
  \bibinfo{author}{P.~Ferreira}, \bibinfo{author}{S.~Kapoor},
  \bibinfo{author}{S.~Vanka},
\newblock \bibinfo{title}{Finite volume simulation framework for die casting
  with uncertainty quantification},
\newblock \bibinfo{journal}{Applied Mathematical Modelling}
  \bibinfo{volume}{74} (\bibinfo{year}{2019}) \bibinfo{pages}{132--150}.
\bibitem[{Shahane(2019)}]{shahane2019numerical}
\bibinfo{author}{S.~Shahane}, \bibinfo{title}{Numerical simulations of die
  casting with uncertainty quantification and optimization using neural
  networks}, Ph.D. thesis, University of Illinois at Urbana-Champaign,
  \bibinfo{year}{2019}.
\bibitem[{Bennon and Incropera(1987)}]{bennon1987continuum}
\bibinfo{author}{W.~Bennon}, \bibinfo{author}{F.~Incropera},
\newblock \bibinfo{title}{A continuum model for momentum, heat and species
  transport in binary solid-liquid phase change systems—i. model
  formulation},
\newblock \bibinfo{journal}{International Journal of Heat and Mass Transfer}
  \bibinfo{volume}{30} (\bibinfo{year}{1987}) \bibinfo{pages}{2161--2170}.
\bibitem[{Plotkowski et~al.(2015)Plotkowski, Fezi, and
  Krane}]{plotkowski2015estimation}
\bibinfo{author}{A.~Plotkowski}, \bibinfo{author}{K.~Fezi},
  \bibinfo{author}{M.~Krane},
\newblock \bibinfo{title}{Estimation of transient heat transfer and fluid flow
  for alloy solidification in a rectangular cavity with an isothermal
  sidewall},
\newblock \bibinfo{journal}{Journal of Fluid Mechanics} \bibinfo{volume}{779}
  (\bibinfo{year}{2015}) \bibinfo{pages}{53--86}.
\bibitem[{Voller and Prakash(1987)}]{voller1987fixed}
\bibinfo{author}{V.~Voller}, \bibinfo{author}{C.~Prakash},
\newblock \bibinfo{title}{A fixed grid numerical modelling methodology for
  convection-diffusion mushy region phase-change problems},
\newblock \bibinfo{journal}{International Journal of Heat and Mass Transfer}
  \bibinfo{volume}{30} (\bibinfo{year}{1987}) \bibinfo{pages}{1709--1719}.
\bibitem[{Bartrons et~al.(2019)Bartrons, Galione, and
  P{\'e}rez-Segarra}]{bartrons2019fixed}
\bibinfo{author}{E.~Bartrons}, \bibinfo{author}{P.~Galione},
  \bibinfo{author}{C.~P{\'e}rez-Segarra},
\newblock \bibinfo{title}{Fixed grid numerical modelling of frost growth and
  densification},
\newblock \bibinfo{journal}{International Journal of Heat and Mass Transfer}
  \bibinfo{volume}{130} (\bibinfo{year}{2019}) \bibinfo{pages}{215--229}.
\bibitem[{Yue et~al.(2018)Yue, Liu, and Wang}]{yue2018freezing}
\bibinfo{author}{X.~Yue}, \bibinfo{author}{W.~Liu}, \bibinfo{author}{Y.~Wang},
\newblock \bibinfo{title}{Freezing delay, frost accumulation and droplets
  condensation properties of micro-or hierarchically-structured silicon
  surfaces},
\newblock \bibinfo{journal}{International Journal of Heat and Mass Transfer}
  \bibinfo{volume}{126} (\bibinfo{year}{2018}) \bibinfo{pages}{442--451}.
\bibitem[{El~Cheikh(2014)}]{el2014effect}
\bibinfo{author}{A.~El~Cheikh}, \bibinfo{title}{The effect of surface
  wettability on frost growth and densification on flat plates}, Ph.D. thesis,
  University of Illinois at Urbana-Champaign, \bibinfo{year}{2014}.
\bibitem[{Huang et~al.(2011)Huang, Liu, Liu, and Gou}]{huang2011preparation}
\bibinfo{author}{L.~Huang}, \bibinfo{author}{Z.~Liu}, \bibinfo{author}{Y.~Liu},
  \bibinfo{author}{Y.~Gou},
\newblock \bibinfo{title}{Preparation and anti-frosting performance of
  super-hydrophobic surface based on copper foil},
\newblock \bibinfo{journal}{International Journal of Thermal Sciences}
  \bibinfo{volume}{50} (\bibinfo{year}{2011}) \bibinfo{pages}{432--439}.
\bibitem[{Liu et~al.(2008)Liu, Gou, Wang, and Cheng}]{liu2008frost}
\bibinfo{author}{Z.~Liu}, \bibinfo{author}{Y.~Gou}, \bibinfo{author}{J.~Wang},
  \bibinfo{author}{S.~Cheng},
\newblock \bibinfo{title}{Frost formation on a super-hydrophobic surface under
  natural convection conditions},
\newblock \bibinfo{journal}{International Journal of Heat and Mass Transfer}
  \bibinfo{volume}{51} (\bibinfo{year}{2008}) \bibinfo{pages}{5975--5982}.
\bibitem[{Sommers et~al.(2018)Sommers, Gebhart, and Hermes}]{sommers2018role}
\bibinfo{author}{A.~Sommers}, \bibinfo{author}{C.~Gebhart},
  \bibinfo{author}{C.~Hermes},
\newblock \bibinfo{title}{The role of surface wettability on natural convection
  frosting: Frost growth data and a new correlation for hydrophilic and
  hydrophobic surfaces},
\newblock \bibinfo{journal}{International Journal of Heat and Mass Transfer}
  \bibinfo{volume}{122} (\bibinfo{year}{2018}) \bibinfo{pages}{78--88}.
\bibitem[{Cai et~al.(2011)Cai, Wang, Hou, and Zhang}]{cai2011study}
\bibinfo{author}{L.~Cai}, \bibinfo{author}{R.~Wang}, \bibinfo{author}{P.~Hou},
  \bibinfo{author}{X.~Zhang},
\newblock \bibinfo{title}{Study on restraining frost growth at initial stage by
  hydrophobic coating and hygroscopic coating},
\newblock \bibinfo{journal}{Energy and Buildings} \bibinfo{volume}{43}
  (\bibinfo{year}{2011}) \bibinfo{pages}{1159--1163}.
\bibitem[{Wang et~al.(2015)Wang, Liang, Yang, Fan, and Zhang}]{wang2015effects}
\bibinfo{author}{F.~Wang}, \bibinfo{author}{C.~Liang},
  \bibinfo{author}{M.~Yang}, \bibinfo{author}{C.~Fan},
  \bibinfo{author}{X.~Zhang},
\newblock \bibinfo{title}{Effects of surface characteristic on frosting and
  defrosting behaviors of fin-tube heat exchangers},
\newblock \bibinfo{journal}{Applied Thermal Engineering} \bibinfo{volume}{75}
  (\bibinfo{year}{2015}) \bibinfo{pages}{1126--1132}.
\bibitem[{Sommers et~al.(2016)Sommers, Truster, Napora, Riechman, and
  Caraballo}]{sommers2016densification}
\bibinfo{author}{A.~Sommers}, \bibinfo{author}{N.~Truster},
  \bibinfo{author}{A.~Napora}, \bibinfo{author}{A.~Riechman},
  \bibinfo{author}{E.~Caraballo},
\newblock \bibinfo{title}{Densification of frost on hydrophilic and hydrophobic
  substrates--examining the effect of surface wettability},
\newblock \bibinfo{journal}{Experimental Thermal and Fluid Science}
  \bibinfo{volume}{75} (\bibinfo{year}{2016}) \bibinfo{pages}{25--34}.
\bibitem[{Liu et~al.(2017)Liu, Zhu, Liu, Jiang, Song, Francisco, Zeng, and
  Wang}]{liu2017distinct}
\bibinfo{author}{J.~Liu}, \bibinfo{author}{C.~Zhu}, \bibinfo{author}{K.~Liu},
  \bibinfo{author}{Y.~Jiang}, \bibinfo{author}{Y.~Song},
  \bibinfo{author}{J.~Francisco}, \bibinfo{author}{X.~Zeng},
  \bibinfo{author}{J.~Wang},
\newblock \bibinfo{title}{Distinct ice patterns on solid surfaces with various
  wettabilities},
\newblock \bibinfo{journal}{Proceedings of the National Academy of Sciences}
  \bibinfo{volume}{114} (\bibinfo{year}{2017}) \bibinfo{pages}{11285--11290}.
\bibitem[{Volmer(1926)}]{volmer1926nucleus}
\bibinfo{author}{M.~Volmer},
\newblock \bibinfo{title}{Nucleus formation in supersaturated systems},
\newblock \bibinfo{journal}{Z. Phys. Chem.} \bibinfo{volume}{119}
  (\bibinfo{year}{1926}) \bibinfo{pages}{277--301}.
\bibitem[{W{\"o}lk and Strey(2001)}]{wolk2001homogeneous}
\bibinfo{author}{J.~W{\"o}lk}, \bibinfo{author}{R.~Strey},
\newblock \bibinfo{title}{Homogeneous nucleation of h2o and d2o in comparison:
  the isotope effect},
\newblock \bibinfo{journal}{The Journal of Physical Chemistry B}
  \bibinfo{volume}{105} (\bibinfo{year}{2001}) \bibinfo{pages}{11683--11701}.
\bibitem[{Becker and D{\"o}ring(1935)}]{becker1935kinetische}
\bibinfo{author}{R.~Becker}, \bibinfo{author}{W.~D{\"o}ring},
\newblock \bibinfo{title}{Kinetische behandlung der keimbildung in
  {\"u}bers{\"a}ttigten d{\"a}mpfen},
\newblock \bibinfo{journal}{Annalen der Physik} \bibinfo{volume}{416}
  (\bibinfo{year}{1935}) \bibinfo{pages}{719--752}.
\bibitem[{Volmer(1939)}]{volmer1939kinetik}
\bibinfo{author}{M.~Volmer},
\newblock \bibinfo{title}{Kinetik der phasenbildung}  (\bibinfo{year}{1939}).
\bibitem[{Twomey(1959)}]{twomey1959experimental}
\bibinfo{author}{S.~Twomey},
\newblock \bibinfo{title}{Experimental test of the volmer theory of
  heterogeneous nucleation},
\newblock \bibinfo{journal}{The Journal of Chemical Physics}
  \bibinfo{volume}{30} (\bibinfo{year}{1959}) \bibinfo{pages}{941--943}.
\bibitem[{Xu et~al.(2015)Xu, Lan, Peng, Wen, and Ma}]{xu2015heterogeneous}
\bibinfo{author}{W.~Xu}, \bibinfo{author}{Z.~Lan}, \bibinfo{author}{B.~Peng},
  \bibinfo{author}{R.~Wen}, \bibinfo{author}{X.~Ma},
\newblock \bibinfo{title}{Heterogeneous nucleation capability of conical
  microstructures for water droplets},
\newblock \bibinfo{journal}{RSC Advances} \bibinfo{volume}{5}
  (\bibinfo{year}{2015}) \bibinfo{pages}{812--818}.
\bibitem[{Varanasi et~al.(2009)Varanasi, Hsu, Bhate, Yang, and
  Deng}]{varanasi2009spatial}
\bibinfo{author}{K.~Varanasi}, \bibinfo{author}{M.~Hsu},
  \bibinfo{author}{N.~Bhate}, \bibinfo{author}{W.~Yang},
  \bibinfo{author}{T.~Deng},
\newblock \bibinfo{title}{Spatial control in the heterogeneous nucleation of
  water},
\newblock \bibinfo{journal}{Applied Physics Letters} \bibinfo{volume}{95}
  (\bibinfo{year}{2009}) \bibinfo{pages}{094101}.
\bibitem[{Harlow and Amsden(1975)}]{harlow1975numerical}
\bibinfo{author}{F.~Harlow}, \bibinfo{author}{A.~Amsden},
\newblock \bibinfo{title}{Numerical calculation of multiphase fluid flow},
\newblock \bibinfo{journal}{Journal of Computational Physics}
  \bibinfo{volume}{17} (\bibinfo{year}{1975}) \bibinfo{pages}{19--52}.
\bibitem[{Le~Gall et~al.(1997)Le~Gall, Grillot, and Jallut}]{le1997modelling}
\bibinfo{author}{R.~Le~Gall}, \bibinfo{author}{J.~Grillot},
  \bibinfo{author}{C.~Jallut},
\newblock \bibinfo{title}{Modelling of frost growth and densification},
\newblock \bibinfo{journal}{International Journal of Heat and Mass Transfer}
  \bibinfo{volume}{40} (\bibinfo{year}{1997}) \bibinfo{pages}{3177--3187}.
\bibitem[{Bartrons et~al.(2018)Bartrons, Oliet, Guti{\'e}rrez, Naseri, and
  P{\'e}rez-Segarra}]{bartrons2018finite}
\bibinfo{author}{E.~Bartrons}, \bibinfo{author}{C.~Oliet},
  \bibinfo{author}{E.~Guti{\'e}rrez}, \bibinfo{author}{A.~Naseri},
  \bibinfo{author}{C.~P{\'e}rez-Segarra},
\newblock \bibinfo{title}{A finite volume method to solve the frost growth
  using dynamic meshes},
\newblock \bibinfo{journal}{International Journal of Heat and Mass Transfer}
  \bibinfo{volume}{124} (\bibinfo{year}{2018}) \bibinfo{pages}{615--628}.
\bibitem[{Eckert and Drake(????)}]{eckertanalysis}
\bibinfo{author}{E.~Eckert}, \bibinfo{author}{R.~Drake},
\newblock \bibinfo{title}{Analysis of heat and mass transfer, mcgraw-hill, new
  york, 1972}  (????).
\bibitem[{Lide and Kehiaian(1994)}]{lide1994crc}
\bibinfo{author}{D.~Lide}, \bibinfo{author}{H.~Kehiaian}, \bibinfo{title}{CRC
  handbook of thermophysical and thermochemical data}, \bibinfo{publisher}{CRC
  Press}, \bibinfo{year}{1994}.
\bibitem[{Fessler(1979)}]{fessler1979wetair}
\bibinfo{author}{T.~Fessler},
\newblock \bibinfo{title}{Wetair: A computer code for calculating thermodynamic
  and transport properties of air-water mixtures}  (\bibinfo{year}{1979}).
\bibitem[{Studnikov(1970)}]{studnikov1970viscosity}
\bibinfo{author}{E.~Studnikov},
\newblock \bibinfo{title}{The viscosity of moist air},
\newblock \bibinfo{journal}{Journal of Engineering Physics and Thermophysics}
  \bibinfo{volume}{19} (\bibinfo{year}{1970}) \bibinfo{pages}{1036--1037}.
\bibitem[{Na and Webb(2004)}]{na2004new}
\bibinfo{author}{B.~Na}, \bibinfo{author}{R.~Webb},
\newblock \bibinfo{title}{New model for frost growth rate},
\newblock \bibinfo{journal}{International Journal of Heat and Mass Transfer}
  \bibinfo{volume}{47} (\bibinfo{year}{2004}) \bibinfo{pages}{925--936}.
\bibitem[{Sanders(1974)}]{sanders1974influence}
\bibinfo{author}{C.~T. Sanders},
\newblock \bibinfo{title}{The influence of frost formation and defrosting on
  the performance of air coolers}  (\bibinfo{year}{1974}).
\bibitem[{{\c{C}}engel(2008)}]{ccengel2008introduction}
\bibinfo{author}{Y.~A. {\c{C}}engel},
\newblock \bibinfo{title}{Introduction to thermodynamics and heat transfer},
\newblock \bibinfo{journal}{McGraw-Hill}  (\bibinfo{year}{2008}).
\bibitem[{Teske et~al.(2005)Teske, Vogel, and Bich}]{teske2005viscosity}
\bibinfo{author}{V.~Teske}, \bibinfo{author}{E.~Vogel},
  \bibinfo{author}{E.~Bich},
\newblock \bibinfo{title}{Viscosity measurements on water vapor and their
  evaluation},
\newblock \bibinfo{journal}{Journal of Chemical \& Engineering Data}
  \bibinfo{volume}{50} (\bibinfo{year}{2005}) \bibinfo{pages}{2082--2087}.
\bibitem[{Harvey(2016)}]{harvey2016properties}
\bibinfo{author}{A.~Harvey},
\newblock \bibinfo{title}{Properties of ice and supercooled water},
\newblock \bibinfo{journal}{CRC Handbook of Chemistry and Physics 97$^{th}$
  Edition}  (\bibinfo{year}{2016}) \bibinfo{pages}{6--12}. \URLprefix
  \url{http://www.softouch.on.ca/kb/data/CRC%20Handbook%20of%20Chemistry%20and%20Physics%20-%2097th%20Edition%20(2016).pdf}.
\bibitem[{Stewart(2009)}]{stewart2009physical}
\bibinfo{author}{K.~Stewart},
\newblock \bibinfo{title}{Physical properties of water},
\newblock \bibinfo{journal}{Encyclopedia of Inland Waters}
  (\bibinfo{year}{2009}). \URLprefix
  \url{https://www.sciencedirect.com/science/article/pii/B9780123706263000077}.
\bibitem[{Lemmon(2016)}]{lemmon2016properties}
\bibinfo{author}{E.~Lemmon},
\newblock \bibinfo{title}{Properties of ice and supercooled water},
\newblock \bibinfo{journal}{CRC Handbook of Chemistry and Physics 97$^{th}$
  Edition}  (\bibinfo{year}{2016}) \bibinfo{pages}{6--16}. \URLprefix
  \url{http://www.softouch.on.ca/kb/data/CRC%20Handbook%20of%20Chemistry%20and%20Physics%20-%2097th%20Edition%20(2016).pdf}.
\bibitem[{Lide(2005)}]{lide2005properties}
\bibinfo{author}{D.~Lide},
\newblock \bibinfo{title}{Thermal conductivity of saturated h2o and d2o},
\newblock \bibinfo{journal}{CRC Handbook of Chemistry and Physics, Internet
  Version}  (\bibinfo{year}{2005}) \bibinfo{pages}{6--4}.
\bibitem[{Beysens(2006)}]{beysens2006dew}
\bibinfo{author}{D.~Beysens},
\newblock \bibinfo{title}{Dew nucleation and growth},
\newblock \bibinfo{journal}{Comptes Rendus Physique} \bibinfo{volume}{7}
  (\bibinfo{year}{2006}) \bibinfo{pages}{1082--1100}.
\bibitem[{Liu(2000)}]{liu2000heterogeneous}
\bibinfo{author}{X.~Liu},
\newblock \bibinfo{title}{Heterogeneous nucleation or homogeneous nucleation?},
\newblock \bibinfo{journal}{The Journal of Chemical Physics}
  \bibinfo{volume}{112} (\bibinfo{year}{2000}) \bibinfo{pages}{9949--9955}.
\bibitem[{Iwamatsu(2011)}]{iwamatsu2011heterogeneous}
\bibinfo{author}{M.~Iwamatsu},
\newblock \bibinfo{title}{Heterogeneous critical nucleation on a completely
  wettable substrate},
\newblock \bibinfo{journal}{The Journal of Chemical Physics}
  \bibinfo{volume}{134} (\bibinfo{year}{2011}) \bibinfo{pages}{234709}.
\bibitem[{Gupta and Ghosh(1946)}]{gupta1946report}
\bibinfo{author}{N.~Gupta}, \bibinfo{author}{S.~Ghosh},
\newblock \bibinfo{title}{A report on the wilson cloud chamber and its
  applications in physics},
\newblock \bibinfo{journal}{Reviews of Modern Physics} \bibinfo{volume}{18}
  (\bibinfo{year}{1946}) \bibinfo{pages}{225}.
\bibitem[{Knight(1971)}]{knight1971experiments}
\bibinfo{author}{C.~Knight},
\newblock \bibinfo{title}{Experiments on the contact angle of water on ice},
\newblock \bibinfo{journal}{Philosophical magazine} \bibinfo{volume}{23}
  (\bibinfo{year}{1971}) \bibinfo{pages}{153--165}.
\bibitem[{Harlow and Welch(1965)}]{harlow1965numerical}
\bibinfo{author}{F.~Harlow}, \bibinfo{author}{J.~Welch},
\newblock \bibinfo{title}{Numerical calculation of time-dependent viscous
  incompressible flow of fluid with free surface},
\newblock \bibinfo{journal}{The Physics of Fluids} \bibinfo{volume}{8}
  (\bibinfo{year}{1965}) \bibinfo{pages}{2182--2189}.
\bibitem[{ASHRAE(2001)}]{handbook2001hvac}
\bibinfo{author}{ASHRAE},
\newblock \bibinfo{title}{Hvac fundamentals handbook},
\newblock \bibinfo{journal}{SI Edition}  (\bibinfo{year}{2001})
  \bibinfo{pages}{6.8}.
\bibitem[{Kwon et~al.(2006)Kwon, Lim, Kwon, Koyama, Kim, and
  Kondou}]{kwon2006experimental}
\bibinfo{author}{J.~Kwon}, \bibinfo{author}{H.~Lim}, \bibinfo{author}{Y.~Kwon},
  \bibinfo{author}{S.~Koyama}, \bibinfo{author}{D.~Kim},
  \bibinfo{author}{C.~Kondou},
\newblock \bibinfo{title}{An experimental study on frosting of laminar air flow
  on a cold surface with local cooling},
\newblock \bibinfo{journal}{International Journal of Refrigeration}
  \bibinfo{volume}{29} (\bibinfo{year}{2006}) \bibinfo{pages}{754--760}.
\bibitem[{Geuzaine and Remacle(2009)}]{geuzaine2009gmsh}
\bibinfo{author}{C.~Geuzaine}, \bibinfo{author}{J.~Remacle},
\newblock \bibinfo{title}{Gmsh: A 3-d finite element mesh generator with
  built-in pre-and post-processing facilities},
\newblock \bibinfo{journal}{International Journal for Numerical Methods in
  Engineering} \bibinfo{volume}{79} (\bibinfo{year}{2009})
  \bibinfo{pages}{1309--1331}.

\end{thebibliography}
\end{document}